\documentclass[twocolumn,10pt,aps,pra,superscriptaddress,longbibliography]{revtex4-1}
\usepackage{amssymb}
\usepackage{amsmath}

\usepackage{graphicx}% Include figure files
\usepackage{dcolumn}% Align table columns on decimal point
\usepackage{bm}% bold math
\usepackage{color}
\usepackage{soul}
\usepackage[english]{babel}
\usepackage{epsfig}
\usepackage{amsmath}
\usepackage{amsfonts}
\usepackage{amssymb}
\usepackage{float}
\usepackage{mathrsfs}

\newcommand{\ra}{\rightarrow}

\usepackage{color}
\usepackage{tikz}
\usetikzlibrary{decorations.fractals}
\usepackage{wrapfig}
\usepackage{color}
\usepackage{graphicx,framed}
\usepackage[colorlinks=true, pdfstartview=FitV, linkcolor=blue, citecolor=blue, urlcolor=blue]{hyperref}
\definecolor{shadecolor}{rgb}{0.95, 0.95, 0.86}

\usetikzlibrary{decorations.markings}

\def\blue#1{\textcolor[rgb]{0,0,1}{#1}}

\def\ra{\rightarrow}

\newcommand{\bt}{\beta}

\newcommand{\G}{\Gamma}
\renewcommand{\O}{\Omega}
\renewcommand{\k}{\varkappa}
\renewcommand{\d}{\delta}

\renewcommand{\o}{\omega}
\newcommand{\g}{\gamma}

\renewcommand{\part}{\partial}

\newcommand{\br}{{\mathbb R}}

\newcommand{\bs}{\boldsymbol}

\newtheorem{theorem}{Theorem}[section]
\newtheorem{example}{Example}[section]
\newtheorem{exercise}{Exercise}[section]

\newtheorem{lemma}{Lemma}[section]
\newtheorem{remark}{Remark}[section]

\newtheorem{proposition}{Proposition}[section]
\newtheorem{corollary}{Corollary}[section]

\newtheorem{definition}{Definition}[section]
\def\le{\left}
\def\ri{\right}

\def\C{{\mathbb C}}
\renewcommand\L{\Lambda}
\def\R{{\mathbb R}}

\def\T{{\mathbb T}}

\def\Rscr{\mathcal R}

\def\bt{\begin{theorem}}
\def\et{\end{theorem}}
\def\bc{\begin{corollary}}
\def\ec{\end{corollary}}
\def\bx{\begin{example}\small}
\def\ex{\end{example}}
\def\bxr{\begin{exercise}\small}
\def\exr{\end{exercise}}
\def\bl{\begin{lemma}}
\def\el{\end{lemma}}
\def\bd{\begin{definition}}
\def\ed{\end{definition}}
\def\bp{\begin{proposition}}
\def\ep{\end{proposition}}

\def\br{\begin{remark}}
\def\er{\end{remark}}

\def\be{\begin{equation}}
\def\ee{\end{equation}}
\def\&{\hspace{-15pt}&}
\def\bea{\begin{eqnarray}}
\def\eea{\end{eqnarray}}
\def\beas{\begin{eqnarray*}}
\def\eeas{\end{eqnarray*}}

\def\C{{\mathbb C}}
\def\R{{\mathbb R}}

\def\a{\alpha}

\def\g{\gamma}
\def\gt{\hat\gamma}

\def\m{\mu}

\def\1{{\bf 1}}
\def\r{\rho}
\def\s{ {\sigma}}

\def\th{ {\theta}}

\def\x{\xi}
\def\z{\zeta}

\def\hf{\frac{1}{2}}

\newcommand{\GE}[1]{{\color{red} {#1}}}

\begin{document}

\title{Spectral theory of soliton and breather gases {for} the focusing nonlinear Schr\"odinger equation}

\author{Gennady  El}
\email[Corresponding author : ]{gennady.el@northumbria.ac.uk}
\affiliation{Department of Mathematics, Physics and Electrical Engineering, Northumbria University, Newcastle upon Tyne, United Kingdom}
\author{Alexander Tovbis}
\affiliation{ Department of Mathematics, University of Central Florida, USA}

\begin{abstract}
Solitons and breathers are localized solutions of integrable systems that can be viewed as ``particles'' of complex statistical objects  called soliton and breather gases.  In view of the growing evidence of their ubiquity in fluids and nonlinear optical media these ``integrable'' gases  present  fundamental interest for nonlinear physics. 
We develop  analytical theory of  breather and soliton gases  by considering a special, thermodynamic type limit of the wavenumber-frequency relations for multi-phase  (finite-gap) solutions  of the focusing nonlinear Schr\"odinger  equation. 
This limit is defined by the locus and the critical scaling of the band spectrum of the associated  Zakharov-Shabat operator, and yields  the nonlinear dispersion relations for a spatially homogeneous breather or soliton gas, depending on the presence or absence of the ``background'' Stokes mode. The key quantity of interest  is the density of states defining, in principle,  all spectral and statistical properties of a soliton (breather) gas. The balance of  terms in the nonlinear dispersion relations determines the nature of the gas: from an ideal gas of well separated, non-interacting breathers (solitons) to a special limiting state, which we term  breather (soliton) condensate, and  whose properties are entirely determined by the pairwise interactions between breathes (solitons).  For  a non-homogeneous breather  gas  we derive a full set of kinetic equations describing slow evolution of the density of states and of its carrier wave counterpart. The kinetic equation for soliton gas is recovered by collapsing the Stokes spectral band.   A number of concrete examples of breather and soliton gases are considered, demonstrating efficacy of the developed general theory with broad implications for nonlinear optics, superfluids and oceanography. 
In particular, our work provides the theoretical underpinning for the recently observed remarkable connection of the soliton gas dynamics with the long-term evolution of spontaneous modulational instability. 
\end{abstract}

\maketitle

\section{ Introduction} There is a rapidly growing interest  in the subject of random solutions to integrable nonlinear dispersive equations prompted by Zakharov's paper  ``Turbulence in integrable systems'' \cite{Zakharov:09}.  The unlikely marriage of integrability and randomness within the framework of ``integrable turbulence''   is motivated by  the complexity of many  nonlinear wave phenomena in physical systems that can be successfully modelled by integrable partial differential equations. Despite  integrability of the mathematical model,  physically reasonable results for such systems can often be obtained only in statistical terms (such as probability density function, power spectrum, correlation function, {\it etc.}). This is particularly true for modulationally unstable media, where small random perturbations, inevitably present in any physical system, rapidly grow, leading to disintegration of a constant or slowly varying background and the establishment of a turbulent nonlinear wave wave field exhibiting spontaneous emergence of localized coherent structures such as solitons and breathers \cite{Agafontsev:15}. Applications of integrable turbulence range from oceanography to nonlinear fiber optics and Bose-Einstein condensates. Indeed, recent  observations in ocean waves \cite{costa_soliton_2014, osborne_highly_2019} and laboratory experiments in optical media \cite{Randoux:14, Walczak:15, Randoux:17}  and classical fluids \cite{giovanangeli_soliton_2018, redor_experimental_2019} provide a growing  evidence of ubiquity and pervasiveness of integrable turbulence in physical systems. Due to  complexity of turbulent nonlinear wave fields,  the majority of the existing studies of integrable turbulence involve extensive numerical simulations, while  an analytical development,  vital for the understanding of this fundamental physical phenomenon, is rather limited.    

Our paper develops an analytical theory of an important class of integrable turbulence called soliton gas, and its natural yet nontrivial generalization which we term breather gas, in the framework of the one-dimensional focusing nonlinear Schr\"odinger (fNLS) equation, a canonical model for the description of the envelope dynamics of weakly nonlinear  quasi-monochromatic waves propagating in dispersive, modulationally unstable media when dissipative processes are negligible.

The notion of  soliton gas---an infinite statistical ensemble of interacting solitons---
was first introduced by V. Zakharov \cite{Zakharov:71}  who derived kinetic equation for  a ``rarefied''  gas of KdV solitons  by considering the modification of the soliton velocity due to the position shifts in its pairwise collisions with other solitons in the gas.   The generalization of Zakharov's kinetic equation to the case of the KdV soliton gas of finite density  was obtained by one of the authors in \cite{el_thermodynamic_2003} by considering  the infinite-phase, thermodynamic type limit of the Whitham modulation equations. In \cite{el_thermodynamic_2003} the soliton distribution function  was identified with the {\it density of states}, the fundamental quantity in the spectral theory of random potentials \cite{lifshits_introduction_1988, pastur_spectra_1992}. The finite-gap theory derivation in \cite{el_thermodynamic_2003}  served as a motivation for a more intuitive, physical construction  of  the kinetic equation for a dense soliton gas of the fNLS equation in \cite{GEl:05}. Very recently the kinetic equation having the same structure as  the kinetic equation for soliton gas was derived  in the framework of the ``generalized hydrodynamics'' for quantum many-body integrable systems \cite{doyon_soliton_2018, doyon_geometric_2018, vu_equations_2019}.  These theoretical studies, along with already mentioned observations  in a variety of physical media,  strongly indicate that   soliton gases represent a fundamental object of nonlinear physics, providing a number of intriguing, novel  connections between  soliton theory, dispersive hydrodynamics \cite{Biondini:16} and statistical mechanics. 
In particular, dynamics of a soliton gas in the  fNLS equation recently attracted considerable attention in relation with the description of the nonlinear stage of  spontaneous modulational instability \cite{Agafontsev:15, gelash_bound_2019} and the rogue wave formation \cite{gelash_strongly_2018}. 

Solitons represent spatially localized, decaying at infinity solutions of the fNLS equation. The presence of a non-zero background gives rise to rich families of space-time localized  fNLS solutions called breathers. Thus
the  fNLS soliton gas dynamics in the presence of a non-zero  background can be viewed as  breather gas. 
If the background of a breather is a plane wave the corresponding  fNLS  solution in a general case is the so-called Tajiri-Watanabe breather \cite{tajiri_breather_1998} with the ``standard'' Akhmediev, Kuznetsov-Ma and Peregrine breathers being its particular cases. The background of a breather  can be given by  one of the nonlinear multiphase fNLS solutions, also known as finite-gap potentials \cite{belokolos_algebro-geometric_1994, Osbornebook} (see  \cite{bertola_rogue_2016}  for a description of rogue waves within finite-gap potentials and \cite{chen_rogue_2018, chen_rogue_2019} for  the explicit constructions of fNLS breather  solutions on periodic (elliptic) and 2-phase  background repectively).    

Our paper is concerned with the analytical description of soliton and breather gases using the tools of nonlinear spectral theory, also known as finite-gap theory, which represents an extension of the celebrated inverse scattering transform (IST) method \cite{Zakharov:72} to problems with periodic and quasi-periodic boundary conditions \cite{novikov_theory_1984, belokolos_algebro-geometric_1994}. 
 While the mathematical development of the paper  involves some technical aspects of the finite-gap theory,  the basic ideas behind the proposed construction are very general,  physically transparent and fundamental. In fact, we show that the kinetic theory of breather gas can be viewed as a broad generalization of the well known kinematic wave theory  \cite{whitham} to the case of  nonlinear dispersive random waves described by the fNLS equation. 
 Below we outline the paper organization and  present a high-level description of the main ideas and results. The mathematical underpinnings of  more technical aspects of the paper can be found in the Appendix.
 
First we derive nonlinear dispersion relations for finite-gap potentials of the fNLS equation, which generalize the well known notion of the dispersion relation $\omega=\omega_0(k)$ for linearized waves.  The finite-gap potential $\psi_n(x,t)$ is characterised by $n$-component wavenumber $\bf k$  and frequency $\bs \omega$
 vectors, where $n \in \mathbb{N}$ is the  {\it genus} of the solution. The genus of the solution is equal to the number of nonlinear wave modes or phases comprising  the wave field $\psi(x,t)$ described by the fNLS equation. Within this classification, the plane wave (condensate)  with $|\psi|=1$  represents a genus zero solution.  The well known elliptic solutions of the fNLS  equation are genus one solutions, with the standard (fundamental) solitons being a degenerate case of the genus one.  The standard breathers (Akhmediev, Kuznetsov-Ma and Peregrine)  all represent degenerate genus two solutions \cite{Osbornebook}.  In our construction of breather gas  we  assume an even genus $n=2N$; the results for the potentials of an odd genus (yielding  soliton gas in the appropriate limit) are obtained by ``collapsing'' the breather gas background to zero.
  
Nonlinear dispersion relations for the fNLS $n$-gap potentials represent a system of linear equations \eqref{WFR}, \eqref{WFR-tilde} relating ${\bf k}$ and $\bs \omega$ with other parameters of the solution, which are most conveniently expressed in terms of the band spectrum $\Sigma_n \in \mathbb{C}$ of the Zakharov-Shabat (ZS) operator associated with the fNLS equation \cite{Zakharov:72}. We can symbolically represent these relations in a parametric form   
\be\label{n_gap_disp}
{\bs \omega}= {\bs \Omega} (\Sigma_n), \qquad {\bf k} = {\bf K}(\Sigma_n).
\ee

The core of the paper is the derivation  and analysis of the nonlinear dispersion relations \eqref{dr_breather_gen1}, \eqref{dr_breather_gen2} for a breather/soliton gas  which are obtained from relations \eqref{n_gap_disp} by applying a  special  infinite-genus, thermodynamic type limit.  The crucial role in our analysis  is played by:  (i)  a special choice of the wavenumber-frequency set in \eqref{n_gap_disp}, and (ii) the critical, $n$-dependent scalings of the band/gap distributions in the finite-gap potentials. We distinguish between three such scalings: exponential, super-exponential and sub-exponential, each corresponding to a distinct type of  breather/soliton gas. Each type of the scaling implies a specific balance of terms in the nonlinear dispersion relations \eqref{n_gap_disp}, resulting in certain distinct properties of the corresponding soliton/breather gases. 

We show that the super-exponential scaling corresponds to an ``ideal gas'' of  non-interacting, isolated quasiparticles (breathers or solitons), whose dynamics is determined by secular (non-integral) terms in the dispersion relations. In the opposite case of the sub-exponential spectral scaling the  properties of the  gas are entirely determined by the integral, interaction terms, and  the individual characterization of quasiparticles is suppressed. We call the corresponding gas a breather (soliton) condensate. {We show  that a particular case of the  condensate representing a critically dense bound state soliton gas is characterized by a special  density of states distribution \eqref{Weyl}, which coincides with the appropriately normalized semi-classical distribution of  discrete spectrum in the ZS scattering problem for a rectangular potential \cite{Zakharov:72, cohen_solutions_1992,  Jenkins:14}}.  We also present a nontrivial example of a non bound state soliton condensate characterized by a circular  spectral locus in the complex plane, leading to the group velocity of quasi-particles in the condensate being twice the speed of  free solitons for the same spectral parameter. Finally, the  exponential spectral  scaling corresponds to the general case of a soliton/breather gas of finite density, in which the effects related to the individual motion of quasiparticles and their pairwise interactions are in balance.

As a straightforward consequence of the nonlinear dispersion relations of breather/soliton gas in   we derive  an integral equation \eqref{eq_state} relating the  velocity $s(\eta)$ of quasiparticles in a breather/soliton gas (we call them the `tracer' breathers or solitons) with the fundamental quantity $u(\eta)$  called the density of states  \citep{lifshits_introduction_1988, pastur_spectra_1992},   $\eta \in \mathbb{C}$ being the ZS spectral parameter. This integral equation specifying $s=\mathcal{F}[u]$, where $\mathcal{F}$ is a functional, can be viewed as the {\it equation of state} of a spatially  homogeneous ({equilibrium}) gas. The equations of state for breather and soliton gases have the same structure but are characterised by different  forms of  both secular (free particle) and integral (pairwise interaction)  terms. In both cases the interaction kernel is determined by the position shift in a breather-breather  and soliton-solton pairwise interactions respectively. We also derive a ``satellite'' system, Eqs.~\eqref{transport_tilde}, \eqref{dr_breather_gas1p},  \eqref{phase_velocity_BG} describing the spectral distribution of the phase velocity $\tilde s(\eta)$ of the carrier wave in a breather/soliton gas with a given density of states $u(\eta)$. 

The evolution of slowly modulated $n$-gap potentials  is known to be described by the so-called Whitham modulation equations \cite{whitham, kamchatnov_nonlinear_2000} representing a system of quasilinear partial differential equations for weak spatiotemporal deformations of the finite-gap ZS spectrum $\Sigma_n(x,t)$ \cite{dafermos_geometry_1986, dubrovin_hydrodynamics_1989}. The modulation system for the fNLS equation necessarily includes  the  $n$-component wave conservation law \cite{whitham, dafermos_geometry_1986, tovbis_semiclassical_2016}
\be \label{wcl0}
{\bf k}_t = {\bs \omega}_{x},
\ee
which should be complemented by the nonlinear dispersion relations \eqref{n_gap_disp}. The application of  the thermodynamic limit to the ``nonlinear kinematic wave system'' \eqref{wcl0}, \eqref{n_gap_disp} results in a  kinetic equation for the density of states $u(\eta, x,t)$ in a spatially non-homogeneous breather/soliton gas. This equation has the form of a transport equation $u_t+(us)_x=0$, complemented by the $x,t$-dependent equation of state    $s(\eta, x, t) = \mathcal F [u(\eta, x,t)]$. Another consequence of the application of the thermodynamic limit to the system \eqref{wcl0}, \eqref{n_gap_disp} is the satellite  transport equation $\tilde u_t + ( \tilde u \tilde s)_x=0$ for the carrier wave wavenumber $\tilde u(\eta, x,t)$ in a breather/soliton gas. 

Finally,  we derive multi-component hydrodynamic reductions of the kinetic equation for breather gas and obtain the solution to a  ``shock tube'' problem consisting of three disparate constant states separated by two propagating contact discontinuities satisfying appropriate Rankine-Hugoniot conditions.

In this work we do not consider particular realizations $ \psi_n(x,t), \  n \gg 1 $, of the nonlinear random wave field in a soliton or breather gas, which would depend on a specific choice of the initial phase vector ${\bs \theta}^{(0)} \in \mathbb{T}^n$ of $\psi_n$  (see Sec.~\ref{sec:FG} for details).
In the  construction of a spatially homogeneous breather/soliton gas it would be natural to assume that the components of the initial phase vector ${\bs \theta}^{(0)}$  are independent random variables with ${\bs \theta}^{(0)}$ being uniformly distributed on $\mathbb{T}^n$.  In the soliton gas limit  $n\to\infty$ the
uniform distribution of phases on $\mathbb{T}^n$ gets replaced by a suitably normalized Poisson
distribution on $\mathbb{R}$, see \cite{el_soliton_2001}.  
The spectral theory  developed in this paper, however,  is not based on any assumptions about the phases of finite-gap solutions
involved.

Due to the fundamental nature of the fNLS equation as a universal mathematical model describing  nonlinear wave processes in a broad range of dispersive media, the theory developed in this paper can find applications in various physical contexts, particularly nonlinear optics and deep water waves, where complex statistical ensembles of breathers or solitons represent ubiquitous phenomena observed in both experimental and natural conditions.

\section{Finite-gap solutions: basic configuration for and nonlinear dispersion relations }
\label{sec:FG}
In this section we shall describe the basic mathematical objects necessary for the derivation of the nonlinear dispersion relations and the equation of state   of the breather gas. 
We consider the fNLS equation in the form
\begin{equation}\label{NLS}
i  \psi_t +  \psi_{xx} +2 |\psi|^2 \psi=0, 
\end{equation}
where $\quad \psi (x,t)$ is a complex wave field. Various families of exact solutions to the fNLS equation are available due to its integrability via the IST \cite{Zakharov:72}. The key step in the integration of the  fNLS equation by the IST is the determination of the spectrum of a linear (Dirac) operator  with the potential $\psi(x,t)$ -- the ZS scattering problem. The fNLS evolution (\ref{NLS})  is characterised by the  ZS spectrum that has a very simple time dependence. 
 
The original IST method enables the construction of fNLS solutions in the class of functions (potentials) sufficiently rapidly decaying at infinity. The long-time asymptotics of such solutions include solitons (discrete ZS spectrum) and linear dispersive waves (continuous spectrum).  Various methods including  Darboux transformation and Hirota's bilinear method  enable the construction of  exact solutions describing solitons on finite background, also known as  breathers (see e.g. \cite{ablowitz_homoclinic_1990, tajiri_breather_1998, ohta_general_2012, akhmediev_rogue_2009, bilman_robust_2019}).

An extension of  the IST to a certain class of periodic and quasi-periodic potentials,  the so-called finite-gap theory (FGT)  \cite{novikov_theory_1984, belokolos_algebro-geometric_1994}, enables the construction of  a broad range of non-decaying NLS solutions, which include solitons and breathers as some particular, limiting cases. The ZS spectrum of finite-gap solutions consists of a finite Schwarz-symmetric collection of curvilinear segments  $\g_j\subset\C$  called bands. Here Schwarz symmetry means  that if $z\in \C$
 is  a  point of the spectrum then so is the c.c. point $\overline z$.

An   $n$-gap solution $\psi=\psi_n(x,t)$ of \eqref{NLS} is defined by a fixed set of  $2(n+1)$ endpoints of spectral bands  $\g_j$, $j=1, \dots, n+1$,  and depends on $n$ real phases 
$\bs{\theta}(x,t)=\bs{k} x+\bs{\o} t +\bs{\theta}^0$ with the initial phase vector $\bs{\theta}^0  \in \T^n$, so that  $|\psi_n(x,t)| = F_n(\bs{\theta}(x,t))$, where $F_n$ is a multi-phase (quasiperiodic) function in both $x$ and $t$, that can be expressed in terms  of the Riemann Theta-functions,  \cite{its_explicit_1976}. The $n$-component wavenumber $\bs k $  and the frequency $\bs \o $ vectors depend on the endpoints $\{\a_{ j} ,  j= 0, 1, 2, \dots, n\} $ of the spectral bands, which define 
a hyperelliptic Riemann surface $\Rscr$ of genus $n$ 
given by: 
\begin{equation}
\label{rsurf}
R(z)=\prod_{j=0}^{n}(z-\a_j)^\hf(z-\bar\a_j)^\hf,\quad
\alpha_j = a_j + i b_j, \  \ b_j>0,  \end{equation}
$z \in \C$ being a complex spectral parameter in the ZS scattering problem; $\blue{R(z)} \sim z^{2n+1}$ as $z \to \infty$.  The branch cuts of $R(z)$ will be specified below.
{Finite-gap theory of the 
focusing NLS equation,  originally developed in \cite{its_explicit_1976, ma_periodic_1981},  has been realized in \cite{tracy_nonlinear_1988} as a powerful analytical tool for the understanding of the fundamental phenomenon of modulational instability. Finite-gap NLS solutions have since been used in a number of 
physical applications, notably in water waves (see \cite{Osbornebook} and references therein), and in fiber optics \cite{kamalian_periodic_2016, kamalian_signal_2018}.}
We note here that the finite-gap theory provides a natural framework for the construction of random solutions to the NLS equation by assuming a uniform distribution of the initial phase vector ${\bs  \theta}^{0} \in \T^n$ \cite{osborne_behavior_1993, bertola_rogue_2016, osborne_breather_2019}.
\begin{figure}
\centering
\includegraphics[width= 1 \linewidth]{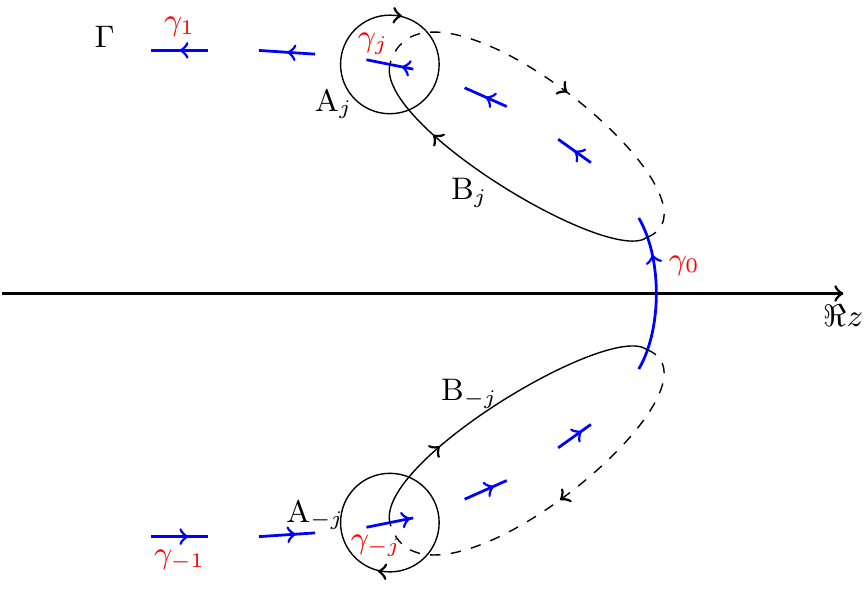}
 \caption{The spectral bands $\g_{\pm j}$ and the basis of cycles $\mathrm{A}_{\pm j}, \mathrm{B}_{\pm j}.$
 The 1D curve $\G$ consists of the bands $\g_j$ and gaps between the bands.}
 \label{Fig:Cont}
\end{figure}

We initially assume that  all spectral bands lie along a finite, 1D Schwarz-symmetric curve $\G$ (see Fig.~\ref{Fig:Cont}); later this restriction will be removed  to allow the bands to be located in a 2D compact subset of $\C$. {The order
we use to enumerate the bands along $\G$ is shown on Fig.~\ref{Fig:Cont}. 
The curve $\G$ does not have to be  a connected curve; the exact meaning of this statement will become clear in Section \ref{sec-therm}.}
We also assume that  the genus $n$ is even, $n=2N$, $N \in \mathbb{N}$; the transition to an odd genus case will be described below.   We enumerate the branch points according to the following notations for the spectral bands. 
 The spectral band $\g_j$,  $j=1,\dots,N$,   is defined as the segment of $\G$ in the upper half plane $\C^+$  connecting the branch points $\a_{2j}$ and $\a_{2j+1}$;
 the spectral band $\g_{-j}$ is Schwarz symmetric to $\g_j$ --- it connects, naturally, the  branch points $\bar\a_{2j}$ and $\bar\a_{2j+1}$ which,
 for convenience, we will denote as $\a_{-2j}$ and $\a_{-2j-1}$ respectively. Finally, 
 there is  an {\it exceptional band} band  $\g_0$ crossing the real axis and connecting  the branch points $\a_{1}$ and $\a_{-1}=\bar\a_1$. An odd genus case can then formally be considered by collapsing 
 the  exceptional band into a point on the real axis.
 Note that generally, one can have more than one exceptional band,  $\g_0^1, \g_0^2, \dots \g_0^m$. We remark that, in the literature, the exceptional bands are sometimes  associated with the so-called Stokes modes \cite{Osbornebook} since, considered in isolation, $\g_0$ represents a ``spectral portrait'' of the plane  wave solution of the fNLS equation, also called a Stokes wave.   
 
 We can now define spectral gaps $c_j, j=1, \dots N$  as segments of $\Gamma$ connecting the endpoint $\a_{2j+1}$ of  $\g_j$ with the 
endpoint $\a_{2j+2}$ of the neighboring band $\g_{j+1}$ (or $\a_1$ of the exceptional band $\g_0$ if $j=N$), 
 and their c.c. (with $j<0$).  

We define two sets of contours  on the Riemann surface $\Rscr$: the  $\mathrm{A}_j$-contours  surrounding the bands $\g_j$  clockwise, $|j|=0,\cdots, N$, and  the  $\mathrm{B}_j$-contours, $j=1,\cdots, N$, so that  $\mathrm{B}_j$ starts at the exceptional  band $\g_0$,  goes to the  band $\g_j$  on the upper sheet clockwise and returns to the  band  $\g_0$ on the lower sheet of the Riemann surface.  
For $j<0$ we define $\mathrm{B}_j=\bar {\mathrm{B}}_{-j}$ {with the same clockwise orientation}, see Fig.~\ref{Fig:Cont}.

We notice that the contours $\mathrm{A}_j$, $|j|=1,\cdots,N$, and  $-\mathrm{B}_j$,  
$\mathrm{B}_{-j}$,  $j=1,\dots,N$,    form a homology basis  of the Riemann surface $\Rscr$ with
the branchcuts along $\g_j$,  $|j|=0,1,\dots,N$.

\begin{figure}
\centering
\includegraphics[width= 1 \linewidth]{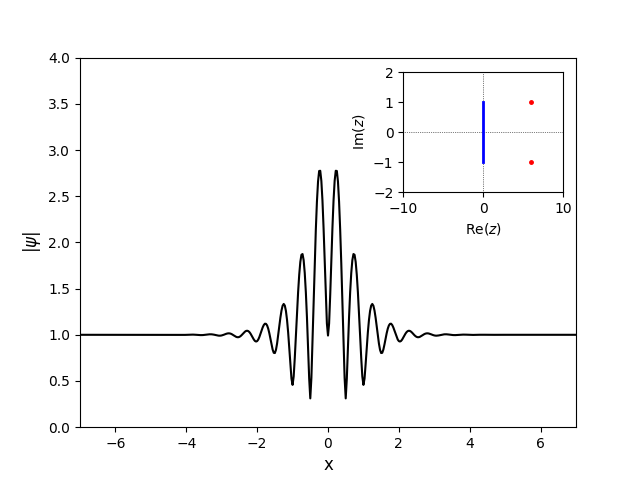}
 \caption{Tajiri-Watanabe (TW) breather (a soliton on finite background). Shown is $|\psi_{TW}(x,0)|$ and the  spectral portrait (inset). The spectral parameters are:  $\a_1=i$,  $\a_2 = 6+0.996i$.}
 \label{fig:TW}
\end{figure}

Collapsing a single pair of bands $\g_{\pm j}$, $j > 0$ into a pair of double points ($\a_{\pm(2j+1)} \to \a_{\pm 2j}$) implies the appearance of a soliton on a $(n-1)$-gap solution background. Shrinking all bands $\g_{\pm j}$, $j=0,1, \dots, N$ to points  corresponds to the transformation of a $2N$-gap solution into a $2N$-soliton solution with $b_j$ corresponding to soliton amplitudes and $-4a_j$ -- to their velocities at $t \to \infty$ (cf. Eq. \eqref{rsurf} for the definitions of $a_j, b_j$) provided that all $a_j$ are distinct. If all $a_j$ are equal (without loss of generality one can assume that $a_j=0 \ \forall j$, i.e.  each $\g_j$ is an interval on the imaginary axis) the limiting $N$-soliton  solution is a {\it bound state} \cite{Zakharov:72, li_degenerate_2017}, in which case all
the solitons do not separate as $t \to \infty$.  We note that in the soliton limit the exceptional band $\g_0$  collapses to the origin (the collapsed band can be  associated with a zero-amplitude soliton) and so does not contribute to the limiting $N$-soliton solution. If the exceptional band $\g_0$ remains finite, then collapsing all the other bands into double points correspond to a $N$-breather limit of a $2N$-gap solution with the finite band  $\g_0$ being ``responsible'' for the background (indeed, the genus zero solution with a single band $\g_0=[-iq,iq]$, $q>0$, in the spectrum is the so-called plane wave, or the condensate, given by $\psi = q e^{i 2 q^2 t}$).  The generic ``elementary'' breather corresponding to a degenerate genus $2$ solution is the so-called Tajiri-Watanabe (TW) breather \cite{tajiri_breather_1998} with the typical behavior of the amplitude $|\psi_{TW}(x,t)|$ shown in Fig.~\ref{fig:TW}. The spectral portrait of the TW breather (shown in the inset of Fig.~\ref{fig:TW}) consists of  the vertical band connecting the points $\pm iq$,  and two double points: $\lambda= a+ib$ and its c.c. (note that $\lambda= \a_2$, $\overline \lambda = \a_{-2}$ within our general finite-gap construction).
The analytical expression for the TW breather solution is available elsewhere (see e.g. \cite{tajiri_breather_1998, slunyaev_nonlinear_2002, gelash_formation_2018}), here we only present its group (envelope) and phase (carrier) velocities: 
\begin{equation}\label{TW_speed}
c_{g}=-2\frac{\Im [\lambda R_0(\lambda)]}{\Im R_0(\lambda)} \equiv s_{TW}(\lambda),  \quad c_p = - \frac{2\Re[\lambda R_0(\lambda)]}{\Re R_0(\lambda)},  
\end{equation} 
where $R_0(\lambda)=\sqrt{\lambda^2 + q^2}$.
All the ``standard'' breathers such as the Akhmediev breather (AB), Kuznetsov-Ma (KM) breather and the Peregrine soliton (PS) playing  prominent role in the rogue wave theories \cite{kharif_rogue_2009, onorato_rogue_2013} are particular cases of the TW breather with $\g_0 = [-i q; i q]$ for some $q>0$, and  the double points $\a_{\pm 2}=\pm ip$, $p>0$, 
with $p<q$ (AB), $p>q$ (KM) and $p=q$ (PS).
The transition from the TW breather solution to the fundamental soliton is achieved by $q \to 0$.  The fNLS fundamental soliton solution is given by \cite{Zakharov:72}
\begin{equation}\label{nls_soliton}
\psi_S (x,t)= 2ib \, \hbox{sech}[2b(x+4at-x_0)]e^{-2i(ax + 2(a^2-b^2)t)+i\phi_0},
   \end{equation}
where $x_0$ is the initial position of the soliton and $\phi_0$ its initial phase. The soliton (envelope) group velocity is $c_g=-4 a= -4 \Re \lambda$ and the carrier phase velocity of a moving ($a \ne 0$) soliton is $c_p=(b^2-a^2)/a = -2\Re (\lambda^2)/\Re \lambda$, in full agreement with the respective TW breather expressions  \eqref{TW_speed} in the limit $q \to 0$.

The wavenumber and frequency vectors, $\bf k$ and $\bs \o$ respectively,  associated with a given finite-gap solution $\psi_{2N}(x,t)$, are not uniquely defined as any  linear combination  of the wavenumber (frequency) vector components with integer coefficients is also a wavenumber (frequency).  Here we introduce two special vectors: ${\bf k} = (k_1, \dots, k_N, \tilde k_1, \dots, \tilde k_N)$ and  ${\bs \o} = (\o_1, \dots, \o_N, \tilde \o_1, \dots, \tilde \o_N)$ whose components are defined as follows  (see Appendix B for details)
\begin{eqnarray}
k_j &=& -\oint_{\mathrm{A}_j}dp, \quad \o_j = -\oint_{\mathrm{A}_j}dq, \quad j=1, \dots, N, \label{kdp}\\
\tilde k_j &=& \oint_{\mathrm{B}_j}dp, \quad \tilde \o_j = \oint_{\mathrm{B}_j}dq, \quad j = 1, \dots, N. \label{omdq}
\end{eqnarray}
The signs of integrals in \eqref{kdp}, \eqref{omdq} will be opposite if we replace $j$ by $-j$. Here  
$dp(z)$ and $dq(z)$ are the meromorphic quasimomentum and quasienergy differentials  with the only poles at $z=\infty$ on both sheets, and defined by (see e.g. \cite{dafermos_geometry_1986}, \cite{tovbis_semiclassical_2016})
\begin{equation}\label{pq}
dp=1 + \mathcal{O}(z^{-2}) , \qquad dq = 4z+\mathcal{O}(z^{-2}) 
\end{equation}
near $z=\infty$ on the main sheet respectively and the normalization conditions requiring that all the periods of $dp,dq$ are real (real normalized
differentials). {The wavenumbers and frequencies can be symmetrically extended to negative indices by $k_{-j}=k_j,\o_{-j}=\o_j$, $j=1, \dots, N, $ 
and similar equations for the ``tilded'' quantities. They also satisfy the corresponding equations \eqref{kdp}, \eqref{omdq}, but 
the signs of integrals in \eqref{kdp}, \eqref{omdq} will be opposite when we replace $j$ by $-j$.}

The proof that $k_j$, $\tilde k_j$, $\o$, $\tilde \o$ defined by \eqref{kdp}, \eqref{omdq} are indeed  wavenumbers and  frequencies of the finite-gap fNLS solution associated with the spectral surface  $\Rscr$ of \eqref{rsurf} can be found in Appendix B.  
We shall call the special set of wavenumbers and frequencies defined by  \eqref{kdp}, \eqref{omdq} {\it the  fundamental wavenumber-frequency set}. 
 We note that the wavenumbers and frequencies defined by  \eqref{kdp} and those defined by \eqref{omdq} are of essentially different nature which is clarified below. 
 
 Let us introduce  two new quantities 
 
 \be\label{eta_delta}
\eta_j=\hf(\a_{2j}+\a_{2j+1}),~~~~\d_j=\hf(\a_{2j}-\a_{2j+1}),
\ee
where $j=1,\dots, N$.  We shall call  the point $\eta_j$  the center of the $j$-th band $\g_j$ and $2|\d_j|$ the $j$-th band width. 
In the lower half plane, we denote $\eta_{-j}=\bar \eta_j$ and $\d_{-j}=\bar \d_j$.  
We also denote the point of intersection of $\G$ with the real axis as $\eta_0$ and the end of the exceptional band $ \a_1$ as $\eta_0+\d_0$. 

 It then follows that the wavenumbers and frequencies defined by  \eqref{kdp} and \eqref{omdq} have drastically different asymptotic properties in the soliton/breather limit, when one of the non-exceptional spectral bands collapses into a double point, $\a_{2i+1}, \a_{2j} \to \eta_j $ (see the end of this section for a qualitative explanation):
\bea
&& \d_j \to 0
  \ \  \implies 
    k_j, \o_j \to 0,  \quad \tilde k_j, \tilde \o_j = \mathcal{O}(1),  \label{breather_lim} \\ 
&& j=1, \dots, N. \nonumber
\eea 
In particular, for  $N=1$ (genus 2), the limit  \eqref{breather_lim} ($k_1 \to 0$, $\omega_1 \to 0$) with a  non-zero band $\g_0$  (i.e.  $\a_1 \ne \overline \a_1$) corresponds to the breather limit 
of the corresponding  two phase nonlinear wave solution. The remaining wavenumber and frequency
$\tilde k_1= \mathcal{O}(1)$,  $\tilde \o_1 = \mathcal{O}(1)$ correspond to the ``carrier'' wave of the TW breather (see Fig.~\ref{fig:TW}). 

 Motivated by these properties for $N=1$ we shall  call the components $k_j,\o_j$  of the wavenumber and the frequency vectors   $\bf k$ and $\bs \o$  the {\it solitonic components} and the components $\tilde k_j$, $\tilde \o_j$---the {\it carrier components}.

It is instructive to characterize the  limiting transitions from the TW breather to the AB, KM and PS in terms of appropriate limits of the fundamental wavenumber-frequency set. This will enable us later  to identify special cases of  breather gas such as, say, PS gas or AB gas. The limiting transitions to the standard breathers are achieved in the following ways (assuming  $\Im \delta_0 \ne 0$):
\begin{equation}\label{TW_to_AB}
\hbox{TW} \to \hbox{AB}: \    \tilde \o_1 \to 0, \   \tilde k_1= \mathcal{O}(1),
\end{equation}
\begin{equation}\label{TW_to_KM}
\hbox{TW} \to \hbox{KM}: \   \tilde \o_1 = \mathcal{O}(1), \   \tilde k_1 \to 0,
\end{equation}
\begin{equation}\label{TW_to_PS}
\hbox{TW} \to \hbox{PS}: \    \tilde \o_1 \to 0, \   \tilde k_1 \to 0.
\end{equation}

The key role in our construction of a breather gas is played by the nonlinear dispersion relations for finite-gap fNLS solutions. In the linear wave theory,  dispersion relation connects the frequency of the linearized mode with its wavenumber. For nonlinear waves, these relations are more complicated, involving other parameters such  mean,  amplitude etc. \cite{whitham}. In the case of integrable equations, the most natural parametrization occurs in terms of the finite-gap spectrum so that the nonlinear dispersion relations have the form ${\bf k}=\bf k({\bs \a})$, ${\bs \omega} = {\bs \omega}({\bs \a}) $ \cite{dubrovin_hydrodynamics_1989}, where the vector $\bs \a$ components are the branch points $\a_j$, 
$|j|=1,\dots, 2N+1$. 

One can show that  for the wavenumbers and frequencies \eqref{kdp}, \eqref{omdq} associated with the finite-gap fNLS solution the  nonlinear dispersion relations are given by (see Appendix C for the proof) 
\bea
&& \tilde k_j+\sum_{|m|=1}^{N} k_m\oint_{\mathrm{B}_m}{{P_j(\z)d\z}\over{R(\z)}}=-\hf \oint_{\gt}{{\z P_j(\z) d\z}\over{R(\z)}}, \nonumber\\
&&\tilde \o_j+\sum_{|m|=1}^{N} \o_m\oint_{\mathrm{B}_m}{{P_j(\z)d\z}\over{R(\z)}}=- \oint_{\gt}{{\z^2 P_j(\z) d\z}\over{R(\z)}}, \nonumber \\
&& |j|= 1,\dots, N,  \label{WUPjM}
\eea
where $\gt$ is a large clockwise oriented contour containing $\G$,
\be\label{Pj}
P_j(z)=\k_{j,1}z^{2N-1}+\k_{j,2}z^{2M-2}+ \dots +\k_{j,2N}
\ee
and $\k_{i,j}$ are the coefficients of the normalized holomorphic differentials $w_j$  defined by: 
\be\label{D-P}
w_j=[P_j(z)/R(z)] dz, \quad \oint_{\mathrm{A_i}}w_j = \delta_{ij}, \quad i, j= \pm 1, \dots, \pm N.
\ee
Taking real and imaginary parts of \eqref{WUPjM} and using the residues in the right hand side,  we obtain separate systems 
for the  solitonic components $k_m$, $\o_m$ \eqref{kdp}:
\bea 
&&\sum_{|m|=1}^{N} k_m\Im\oint_{\mathrm{B}_m}{{P_j(\z)d\z}\over{R(\z)}}=\pi \Re\k_{j,1},\nonumber \\
&&\sum_{|m|=1}^{N} \o_m\Im\oint_{\mathrm{B}_m}{{P_j(\z)d\z} 
\over{R(\z)}}=2\pi \Re(\k_{j,1}\sum_{k=1}^{2N+1}\Re\a_k +\k_{j,2}), \nonumber \\
&& |j|= 1,\dots, N, \label{WFR}
 \eea
 and the carrier components   $\tilde k_m$, $\tilde \o_m$ \eqref{omdq}:
 \bea 
 &&\tilde k_j+\sum_{|m|=1}^{N} k_m\Re\oint_{\mathrm{ B}_m}{{P_j(\z)d\z}\over{R(\z)}}=-\pi \Im\k_{j,1},  \qquad \qquad \nonumber \\
&&\tilde \o_j+\sum_{|m|=1}^{N} \o_m\Re\oint_{\mathrm{ B}_m}{{P_j(\z)d\z}\over{R(\z)}}   \nonumber \\ 
&&= -2\pi \Im(\k_{j,1}\sum_{k=1}^{2N+1}\Re\a_k 
+\k_{j,2}), \nonumber \\ 
&& |j|= 1,\dots, N   \label{WFR-tilde}
\eea
---of the wavenumber and frequency vectors.  In particular, for $N=1$ one can show  that Eqs. \eqref{WFR} and  \eqref{WFR-tilde} imply  that ${\o_1}/{k_1}\ra c_g$ and 
${\tilde\o_1}/{\tilde k_1}\ra c_p$ in the breather limit $\d_1\ra 0$, where $c_g,c_p$ are defined in \eqref{TW_speed}.

{We are now in position to establish the key property \eqref{breather_lim} for the wavenumbers and frequencies in the soliton limit. Indeed, one can see that relations \eqref{WFR} together with  \eqref{ass-coeff} imply  that, for fixed $N$, the solitonic wavenumber and frequency $k_j,\omega_j$ go to zero as the $j$th band width $|\delta_j| \to 0$.  Indeed, in this case,  the integral $\oint_{\mathrm{B}_j} \frac{P_j}{R} dz$ behaves like $\log \delta_j$ due to the contour $\mathrm{B_j}$ crossing the shrinking band $\g_j$ (see Fig.~\ref{Fig:Cont}), whereas the remaining integrals (the coefficients
of the linear system) remain bounded. Thus,  to keep balance of terms in \eqref{WFR} $k_j$ and $\omega_j$ must go to zero together with  $\delta_j$,
whereas the carrier wavenumbers $\tilde k_j$ and frequencies $\tilde \omega_j$ given by system \eqref{WFR-tilde} generally remain $\mathcal{O}(1)$. }

We  note that the relations similar to Eqs.~\eqref{WFR}  for the solitonic components of the wavenumber and frequency vectors also arise in the finite-gap KdV theory \cite{flaschka_multiphase_1980}, where they follow from the  relations between real periods and imaginary quasi-periods of the finite-gap KdV solution, with the mapping between the two being realized by the Riemann period matrix. Equations \eqref{WFR-tilde} do not have a KdV counterpart.

\section{Thermodynamic spectral scalings}\label{sec-therm}

\subsection{1D case}\label{therm_1D}
We shall refer to the configuration described in the previous section, when the spectrum of the finite-gap potential is located on a 
Schwarz symmetric curve $\G\subset \C$, as the 1D case. While being quite restrictive, this configuration provides a major insight into the spectral properties of breather and soliton gases in the more  physically realistic  2D case, where  the  bands $\g_j$  are located in some (Schwarz symmetric) region $\L \subset \C$. It also covers the case of a bound state soliton/breather gas, when $\G$ lies on a  vertical line (so that all the solitons  in the gas have the same velocity).

Recall that we assumed an even genus, $n=2N$. Due to the symmetry of the curve $\G$ (which may consist of several arcs)  it is sufficient to consider only the upper complex half-plane {($\C^+$)} part  of it, which we denote $\G^+$ {(so that $\G^+=\G\cap\C^+$)}.  We shall be interested in a special, large $N$ limit of the nonlinear dispersion relations \eqref{WFR}.
The main requirements of this limit is that all the gaps {cannot shrink faster than  $\mathcal{O}(N^{-1})$ and all but finitely many
of them are  of the  order  $\mathcal{O}(N^{-1})$. At the same time}  all but finitely many bands are of order much 
smaller than  $\mathcal{O}(N^{-1})$.
We assume that the only bands with the width $\mathcal{O}(1)$ are the exceptional bands, i.e. the Stokes bands crossing the real axis.  In what follows we shall be assuming that there is at most one exceptional band $\g_0$ although our results could be readily extended to the case of finitely many exceptional bands.

{We assume that the shrinking bands fill  all the curve $\G$ except the exceptional band $\g_0$ and the gaps adjacent to it with some density $\varphi>0$. In a more general setting, one can consider the bands $\g_j$, $j\neq 0$,
filling only certain (Schwarz symmetrical) parts of $\G$, separated from each other by exceptional order 1 gaps.
With slight abuse of notations, we will keep the notation $\G$ for the locus of accumulation of the  bands $\g_j$ on 
the original curve $\G$. We also assume $\varphi(\m)|d\m|$ to be a probability measure on $\G^+$.}
Since collapsing  a Schwarz-symmetric pair of (non-exceptional) bands  into a pair of complex conjugate double points corresponds to the appearance of a soliton, 
and the finite exceptional band, considered in isolation, corresponds to the plane wave, 
it is natural to associate the finite-band potential with $N \gg 1$ and all but one bands being close to ``collapse'' with the gas of solitons on a finite background, i.e. breather gas. In the case of no exceptional bands we will have the soliton gas limit,  and in  case of more than one exceptional bands---the generalized breather gas limit. {The generalized breather gas can be viewed as a gas of solitons on the background of $n$-gap fNLS solution with $n \ge 1$. Indeed, multiple exceptional bands, considered in
isolation, would represent the corresponding finite-gap solutions. Considered
together with collapsing (in the appropriate $n \to \infty$ limit) bands, one obtains soliton
gas on the finite-gap background.}

We  now assume that for $N \gg 1$ the centers  
$\eta_j$ of the bands $\g_j$, $j=1,\dots,N,$ are distributed  along  
with some limiting density $\varphi(\mu)>0$,  $\mu \in\G^+$,
that is smooth on  $\G^+$.  It then follows that $|\eta_j - \eta_{j+1}|  \sim 1/N$.

 As for the scaling of the band widths, we consider the following options: 
 \begin{itemize}
  \item 
 [(i)] {\it exponential spectral scaling}: the band widths $2|\d_j|$  of $\g_j$ are exponentially narrow in $N$:
 \be\label{exp_scal}
 |\d_j|\sim e^{-N \tau{(\eta_j)}}, \quad |j|= 1, \dots, N,
 \ee
 where $\tau(\mu)$ is a smooth {positive}
 function on $\G^+$ having the meaning of the normalised logarithmic  band width ($\tau(\eta_j) \sim -\ln |\d_j|/N)$. The limit obtained in this scaling will be referred to as a (regular) soliton or a breather gas limit depending on the size of the exceptional band (in the soliton gas limit $\delta_0 \to 0$). 
  \item 
 [(ii)] 
 {\it sub-exponential spectral scaling}: for any $a>0$
\be\label{sub-exp_scal}
e^{- a N } \ll  |\d_j|  \ll \frac{1}{N}, \quad |j|= 1, \dots, N,
 \ee
 We shall refer to the  $N \to \infty$ limit obtained in this scaling as a soliton (breather) condensate limit (the reasons for this name will become clear later). It is clear that in this limit $\tau(\eta) \to 0$.
\item 
 [(iii)] 
 {\it super-exponential spectral scaling}: for any $a>0$
\be \label{sup-exp_scal}
e^{-a N } \gg  |\d_j|,  \quad |j|= 1, \dots, N. 
\ee
The limit obtained in this scaling will  be referred to as an ideal  breather or soliton gas limit. In this limit $\tau(\eta) \to \infty$. 
\end{itemize}
Of course, one can also consider the case of simultaneous different scalings on different parts  $\G^+$ (a ``mixed'' scaling). Even though 
such cases can be very interesting,  
we will not discuss them in any detail here.

 Note that in all three scalings $|\delta_j| \ll |\eta_j-\eta_{j+1}|$ so the width of gaps $|c_j| \sim N^{-1}$ and  so $|\d_j|/|c_j| \to 0$ as $N \to \infty$. We then say that in the limit each  collapsed band $\g_j \to \eta_j$ corresponds to a soliton (breather) state within a soliton (breather) gas. Invoking the interpretation of classical solitons as elastically interacting wave-particles, these states can be viewed as quasi-particles that do not necessarily manifest as localized entities in physical space, except in the case of a rarefied  gas.
 We also note that the exponential and sub-exponential  scalings have the ``thermodynamic'' property in the sense that they preserve finiteness of the total density  of waves  $K_N = \sum_{j=1}^N k_{j}$  in the limit $N\to \infty$  so that $\lim_{N \to \infty} K_N= K_{\infty}$,  where $0 < {K_\infty}<\infty$.  Note that for the super-exponential scaling ${K_\infty}=0$.

 \subsection{2D case}\label{sec-2D}
 In the case  when the shrinking  bands $\g_j$, $j>0$ fill a compact 2D region $\L^+$ of the upper complex half-plane,
 the  counterpart of the exponential scaling \eqref{exp_scal} 
 \begin{equation}\label{exp_2D_scal}
 |\d_j|\sim e^{-N^2 \tau{(\eta_j)}}.
 \end{equation}
 Here $\tau(\eta)$ is a positive smooth function
 on $\L^+$. The scaling of the gaps remains 
 $\mathcal{O}(1/N)$, where by the gap width we understand the closest distance  between the bands. 
 In this case  $\varphi(\eta)$ is the 2D density of bands
(and we also distinguish the cases of exponential,
 sub-exponential and super-exponential scalings of bands, similarly to the 1D case). 
 We assume  $\varphi(\eta)>0$ on $\L^+$.
 We shall call such scalings the 2D thermodynamic scalings. In what follows we shall be using the unified notations $\G^+$ for both 1D and 2D configurations, explicitly distinguishing between these cases when necessary.

\bigskip

\section{Nonlinear dispersion relations for breather and soliton gas}

We now proceed with the characterization of  breather  and soliton gases as  thermodynamic limits of finite-gap solutions of fNLS equation. 
For convenience we sometimes will be using the term ``breather gas'' in a generalized sense, assuming that it includes soliton gas as well, the transition to soliton gas being achieved by letting $\delta_0 \to 0$.  We also note that the full construction of a breather gas (which ultimately includes the determination of the random wave field $\psi(x,t)$) implies, along with the thermodynamic spectral scaling,  the random phase approximation, i.e. the uniform distribution of the phase vector ${\bs  \theta}^{(0)} \in \T^n$ \cite{osborne_behavior_1993, el_soliton_2001, bertola_rogue_2016, osborne_breather_2019} but in this paper  we shall be concerned only with the spectral characterization of breather gases. The description of the associated fNLS solutions  (the random process $\psi(x,t)$ generated by the thermodynamic spectral scaling and the uniform phase distribution) will be the subject of a separate work. {We only mention here that the uniform phase distribution on $\mathbb{T}^n$ gets replaced in the thermodynamic limit by the appropriately normalized Poisson distribution on $\mathbb{R}$ as shown in \cite{el_soliton_2001} for the KdV equation.}
\subsection{1D case}
We now apply the 1D thermodynamic spectral scalings to the nonlinear dispersion relations \eqref{WFR}.
Without much loss of generality from now on we shall be assuming that  $\g_0\subset i\R$, i.e. that the exceptional band $\g_0$ lies on the imaginary axis.

We scale the solitonic wavenumbers and frequencies as
\begin{equation}\label{scal_k_nu}
k_j=\frac{\kappa_j}{N}, \qquad \o_j=-\frac{\nu_j}{N}, \qquad N \gg 1,
\end{equation}
so that  $\kappa_j=\kappa(\eta_j)$ and $\nu_j=\nu(\eta_j)$, where  the functions $\kappa(\eta)\geq 0$ {and  $\nu(\eta)$} are smooth interpolations of $\kappa_j,\nu_j$. {We note that  the existence of the interpolating functions $\kappa(\eta)$, $\nu(\eta)$ and the non-negativity of $\kappa(\eta)$ are physically reasonable assumptions that need to be justified mathematically. A clarifying comment will be presented later, when the (integral) equations \eqref{dr_breather_gas1}, \eqref{dr_breather_gas2} for $\kappa(\eta)$ and $\nu(\eta)$  are derived.  We also note that the sign of $\nu(\eta)$ is not fixed.}
 
{The $1/N$ scaling \eqref{scal_k_nu} for the wavenumbers and frequencies 
follows from the requirement  that the diagonal and off-diagonal terms  of  systems \eqref{WFR}  contribute at the same order; this scaling  is  consistent with the exponential rate of shrinking of the bands given Eq.~\eqref{exp_scal} since  $k_j \sim \ln |\delta_j|$ for $|\delta_j| \ll 1$. The scaling different than $1/N$ for $k_j$, $\omega_j$ are possible in the cases  of  sub-exponential or super-exponential spectral distributions given by Eqs.~ \eqref{sub-exp_scal} and \eqref{sup-exp_scal} respectively.
 
 The
functions $\kappa(\eta)$ and $\nu(\eta)$ in \eqref{scal_k_nu} are determined from  $\tau(\eta)$ and
$\varphi(\eta)$, and the geometry of $\Gamma^+$ as we will now explain.}

We denote $R_0(z)=\sqrt{z^2 - \delta_0^2}$ { with the branchcut $[ -\d_0, \d_0]$, where  $\d_0\in i\R^+$ and} the branch of the radical is defined by $R_0(z)\ra z$ as
$z\ra\infty$.
Then, applying the limit $N \to \infty $ to the nonlinear dispersion relations \eqref{WFR} augmented by an exponential spectral scaling 
\eqref{exp_scal} leads to the following  relations  (see Appendix D for  details of the derivation):
\begin{widetext}
\bea\label{dr_breather_gen1}
i \int \limits_{\eta_\infty}^{\eta_1}
\le[\ln \frac{\bar\m-\eta}{\m-\eta}+ \ln \frac{R_0(\eta)R_0(\m)+\eta \m -\d_0^2}
{R_0(\eta)R_0(\bar\m)+\eta \bar\m-\d_0^2}\ri]   u(\m) |d\m|
+i\sigma(\eta)u(\eta) 
=  R_0(\eta)+\tilde u(\eta), \\
i \int \limits_{\eta_\infty}^{\eta_1}
\le[
\ln \frac{\bar\m-\eta}{\m-\eta}+
\ln\frac{R_0(\eta)R_0(\m)+ \eta \m-\d_0^2}
{R_0(\eta)R_0(\bar\m)+ \eta \bar \m-\d_0^2}\ri] v(\mu) |d\m|
+i\sigma(\eta)v(\eta)
= - 2\eta R_0(\eta)+\tilde v(\eta) \label{dr_breather_gen2},
\eea
\end{widetext}
where  $\tilde u$, $\tilde v$ are some smooth functions on $\G^+$ interpolating $\tilde k_j, \tilde \o_j$, that is,
$\tilde u(\eta_j)=\tilde k_j$, $\tilde v(\eta_j)=\tilde \o_j$, $j=1,\dots,N$, 
the integration is performed between $\eta_\infty= \lim \limits_{N \to \infty}\eta_N$ and $\eta_1$ along $\G^+$ and
\begin{equation}\label{dens_states}
u(\eta)= \frac{1}{\pi} \kappa(\eta) \varphi(\eta),\ \  v(\eta)= \frac{1}{\pi}\nu(\eta) \varphi (\eta), \ \ \sigma(\eta)=\frac{2\tau(\eta)}{\varphi(\eta)}.
\end{equation}
Equations \eqref{dr_breather_gen1}, \eqref{dr_breather_gen2} represent general complex 
nonlinear dispersion relations for breather gas. They specify four unknown functions $u(\eta)$, $v(\eta)$, $\tilde u(\eta)$, $\tilde v(\eta)$  in terms of a single nonnegative smooth function $\sigma(\eta)$
and a contour $\G^+$ characterizing  the Riemann surface $\mathcal{R}$ of \eqref{rsurf} in the thermodynamic limit. 
The function $u (\eta) \ge 0$ defined in Eq. \eqref{dens_states} has the meaning of the {\it density of states} \citep{lifshits_introduction_1988, pastur_spectra_1992} as $u(\eta_*) |d \eta|$ is the number of localized 
(soliton or breather) states located in the spectral interval $[\eta_*, \eta_*+ d\eta] \subset \G^+$ and c.c., per unit interval of space (provided  the gas parameters do not depend on $x$). 
{At the intuitive level one can think of the density of states in a soliton gas as of the spectral distribution of solitons ``contained in a portion of gas of unit length" by first assuming the zero boundary conditions at $x=\pm L$, where $L \gg1$ and then normalizing the obtained distribution by $L$. This should be modified for breather gas by replacing zero boundary conditions with the plane wave boundary conditions.}
The integral $ \int_{\eta_\infty}^{\eta_1} u(\eta)|d\eta|$ gives the total integrated density of waves  $K_\infty$ introduced earlier.
The function $v(\eta)$  then represents the temporal counterpart of the density of states. The functions $\tilde u(\eta)$, $\tilde v (\eta)$ can be interpreted as the carrier wavenumber and carrier frequency spectral functions of a breather gas. 
{Note that the integral term in \eqref{dr_breather_gen1}, \eqref{dr_breather_gen2} corresponds to the off-diagonal terms of the 
 linear systems \eqref{WUPjM}, whereas the non-integral (secular) terms in the left hand sides of 
 \eqref{dr_breather_gen1}-\eqref{dr_breather_gen2} correspond to the diagonal terms of \eqref{WUPjM}.
}

Taking the imaginary  part of equations \eqref{dr_breather_gen1}, \eqref{dr_breather_gen2}, we obtain the solitonic component of the breather gas nonlinear dispersion relations 
\begin{widetext}
\bea\label{dr_breather_gas1}
 \int_{\G^+}
\le[\ln\le| \frac{\m-\bar\eta}{\m-\eta}\ri|+ \ln\le|\frac{R_0(\eta)R_0(\m)+\eta \m -\d_0^2}
{R_0(\bar\eta)R_0(\m)+\bar\eta \m-\d_0^2}\ri|\ri]   u(\m) |d\m|
+\sigma(\eta)u(\eta) 
= \Im R_0(\eta), \\
 \int_{\G^+}
\le[\ln\le| \frac{\m-\bar\eta}{\m-\eta}\ri|+\ln\le|\frac{R_0(\eta)R_0(\m)+ \eta \m-\d_0^2}
{R_0(\bar\eta)R_0(\m)+ \bar\eta \m-\d_0^2}\ri|\ri] v(\mu) |d\m|
+\sigma(\eta)v(\eta)
= - 2\Im[\eta R_0(\eta)] \label{dr_breather_gas2},
\eea
\end{widetext}
where, with a slight abuse of notation, we denoted $ \int_{\eta_\infty}^{\eta_1} \dots |d\mu| \equiv \int_{\G^+} \dots |d \mu|$. {We recall here that  the interpolation function $\kappa(\eta)$ was assumed to be non-negative, which, together with positivity of $\varphi(\eta)$, implies non-negativity of the function $u(\eta)=\kappa(\eta)\varphi(\eta)$ defined by Eq.~\eqref{dr_breather_gas1}. 
This is actually a nontrivial fact that, generally speaking, needs to
be proven.  Moreover, the very existence of solutions to the integral equations \eqref{dr_breather_gas1}, \eqref{dr_breather_gas2}, justifying the existence of the interpolating functions $\kappa(\eta)$, $\nu(\eta)$ for the scaled wavenumbers and the frequencies \eqref{scal_k_nu},  is not obvious. In this paper we shall present a number of physically relevant explicit solutions to these equations relegating the general mathematical proofs to an upcoming  publication. }

Taking now real parts of  equations \eqref{dr_breather_gen1}, \eqref{dr_breather_gen2}, we obtain the carrier component of the breather gas dispersion relations (note the choice of the branches of logarithms in \eqref{dr_breather_gas1}, \eqref{dr_breather_gas2} --- see Appendix D, Eq.~\eqref{ass-coeff})
\begin{widetext}
\bea\label{dr_breather_gas1p}
 \int _{\G^+}
\le[\arg \frac{\m-\eta}{\bar\m-\eta}- \arg\frac{R_0(\eta)R_0(\m)+\eta \m -\d_0^2}
{R_0(\eta)R_0(\bar\m)+\eta \bar\m-\d_0^2}\ri]   u(\m) |d\m|
= \Re R_0(\eta) +\tilde u(\eta), \\
\int_{\G^+}
\le[\arg \frac{\m-\eta}{\bar\m-\eta}- \arg\frac{R_0(\eta)R_0(\m)+\eta \m -\d_0^2}
{R_0(\eta)R_0(\bar\m)+\eta \bar\m-\d_0^2}\ri]   v(\m) |d\m|
= - 2\Re[\eta R_0(\eta)]+\tilde v(\eta) \label{dr_breather_gas2p},
\eea
\end{widetext}
which relate the unknown functions $\tilde u(\eta)$, $\tilde v (\eta)$	with the density of states $u(\eta)$ and its temporal analog $v(\eta)$.

In the case  $\d_0\ra 0$, which corresponds to the transition from a breather gas to a soliton gas,  the solitonic dispersion relations \eqref{dr_breather_gas1}, \eqref{dr_breather_gas2} simplify to 
\bea
 \int _{\G^+}\ln \le|\frac{\m-\bar\eta}{\m-\eta}\ri|
u(\m)|d\m|+\sigma(\eta)u(\eta)&= &\Im \eta, \label{dr_soliton_gas1} \\
 \int _{\G^+}\ln \le|\frac{\m-\bar\eta}{\m-\eta}\ri| v(\m)|d\m|+  \sigma(\eta) v(\eta) &=& -4 \Im \eta\Re\eta. \label{dr_soliton_gas2} 
\eea
These are analogs of  similar equations obtained in \citep{el_thermodynamic_2003} for soliton gas of the KdV equation.

The corresponding dispersion relations for the carrier components  become 
\be  \label{dr_soliton_gas1p}
\begin{split}
 \int _{\G^+} \le[\arg \frac{\m-\eta}{\bar\m-\eta} 
-2 \arg \m \ri] u(\m)|d\m| =   \Re \eta + \tilde u(\eta), \\
 \int _{\G^+} \le[\arg \frac{\m-\eta}{\bar\m-\eta}
-2 \arg \m \ri] v(\m)|d\m| =  -2 \Re \eta^2 + \tilde v(\eta).
\end{split}
\ee 
These equations do not have a KdV counterpart as the KdV solitons do not have a carrier component.

The nonlinear dispersion relations \eqref{dr_breather_gas1}  describe the gas of generic moving breathers, which can also be called the TW breather gas. By considering the particular spectral limits  motivated by the limiting cases of the TW breather (Eqs.~\eqref{TW_to_AB} -- \eqref{TW_to_PS}) one can derive the reductions of nonlinear dispersion relations describing gases of   Akhmediev, Kuznetsov-Ma and Peregrine breathers.  A special case of a soliton gas ---the bound state soliton gas --- when all the solitons in the gas move with the same velocity, will be considered in detail later on. 

\subsection{2D case}
We now relax the basic restriction imposed on the spectrum  locus that was used for the derivation of the dispersion relations and the equation of state of the  breather and soliton gases, namely, the requirement that $\eta \in \G$, where $\G$ 
is a 1D Schwarz-symmetric curve in the complex plane. We now assume one of the 2D spectral thermodynamic spectral scalings (exponential, sub-exponential, and super-exponential) when the shrinking  bands $\g_j$ fill a 2D region $\L$ of the complex plane, see  Section III B. The exponential 2D spectral scaling is given by Eq.~\eqref{exp_2D_scal} while the gaps are scaled as $\mathcal{O}(N^{-1})$ (see Section III B).  For the wave numbers and frequencies instead of \eqref{scal_k_nu} we introduce
\begin{equation}\label{scal_k_nu2}
k_j=\frac{\kappa_j}{N^2}, \qquad \o_j=-\frac{\nu_j}{N^2}, \qquad N \gg 1,
\end{equation}
where $\kappa_j=\kappa(\eta_j)$ and $\nu_j=\nu(\eta_j)$, and the interpolating  functions $\kappa(\eta) \geq 0$, $\nu(\eta) \geq 0$ are assumed to be smooth  on $\L^+$.

Then the  2D thermodynamic limit of the nonlinear dispersion relations \eqref{WFR} leads to the same
integral equations \eqref{dr_breather_gas1}-\eqref{eq_state} but with the line integration along $\G^+$ replaced by the integration over a 2D compact domain $\L^+$:
\begin{equation}\label{2D_int}
\int \limits_{\G^+} \dots |d \mu| \to \iint \limits_{\Lambda^+}\dots d \xi d\zeta
\end{equation}
where $\mu = \xi +i \zeta$.
The density of states $u(\eta)$  in the 2D case is defined in such a way that $u(\mu^*)d\xi d\zeta$ gives the number of localized (breather or soliton) states per the element $[\xi^*, \xi^*+d\xi]\times [\zeta^*, \zeta^*+d\zeta]$ of the spectral complex plane and per unit interval of space, assuming spatially uniform  gas.

For convenience  of the exposition, in what follows we shall be using the 1D notation $\int _{\G^+} \dots |d \mu|$ in both 1D and 2D cases
keeping in mind that in {the 2D} case the meaning of the itegral is given by Eq. \eqref{2D_int}.

\section{Equation of state} \label{sec:eq_state}
We now look closer at the solitonic component of dispersion relations  for breather and soliton gases (Eqs.~\eqref{dr_breather_gas1} and ~\eqref{dr_soliton_gas1} respectively).  In both cases elimination of $\sigma(\eta)$  yields  
 a single relation:
\begin{equation}\label{eq_state}
s(\eta) = s_0(\eta) + \int_{\G^+}\Delta(\eta, \mu)[s(\eta) - s(\mu)] u(\mu) |d \mu|,
\end{equation}
where  $s(\eta)=v(\eta)/u(\eta)$, and  $s_0(\eta)$ and $\Delta(\eta, \mu)$ are defined as follows. 

For breather gas: 
\begin{eqnarray}\label{eq_state_breather}
&&s_0(\eta) =  -2  \frac{\Im[\eta R_0(\eta)]}{\Im R_0(\eta)}, \\
 && \Delta(\eta, \mu) =\frac{1}{\Im R_0(\eta)} 
 \le[\ln\le| \frac{\m-\bar\eta}{\m-\eta}\ri|+ \ln\le|\frac{R_0(\eta)R_0(\m)+\eta \m -\d_0^2}
{R_0(\bar\eta)R_0(\m)+\bar\eta \m-\d_0^2}\ri| \ri], \nonumber
\end{eqnarray}
and for soliton gas (obtained from \eqref{eq_state_breather} by letting $\delta_0 \to 0$):
\begin{equation}\label{eq_state_sol}
s_0(\eta) = -4 \Re \eta, \quad \Delta(\eta, \mu) = \frac{1}{ \Im \eta}\ln \le|\frac{\m-\bar\eta}{\m-\eta}\ri|.
\end{equation}

The relation \eqref{eq_state} complemented by  \eqref{eq_state_breather} or  \eqref{eq_state_sol} can be viewed as the {\it equation of state} of a breather (soliton) gas.

Since $\kappa(\eta)$ and $\nu(\eta)$ are the scaled wavenumber and frequency respectively in the thermodynamic limit (see  \eqref{scal_k_nu}), the quantity $s(\eta)= v(\eta)/u(\eta)=\nu(\eta)/ \kappa(\eta)$ in \eqref{eq_state} has a clear physical meaning of the mean  velocity of a ``tracer''  breather (soliton) in a breather (soliton) gas.  As we shall see, in a  spatially inhomogeneous breather gas  the function $s(\eta) \equiv s(\eta, x,t)$ defined by  Eq.~\eqref{eq_state}  with $u(\eta) \equiv  u(\eta; x,t)$ has the meaning of the gas' transport velocity, see Eq.~\eqref{transport} below. For the fNLS equation the soliton gas equation of state  \eqref{eq_state}, \eqref{eq_state_sol} was originally proposed in \cite{GEl:05} using physical reasoning while its KdV analogue had been derived in \cite{el_thermodynamic_2003} using the exponential spectral scaling analogous to \eqref{exp_scal} (as a matter of fact the KdV spectral scaling occurs along the real axis). The derivation presented here provides, along with mathematical justification of the fNLS results of \cite{GEl:05}, their major generalization to the case of breather gas with a number of novel physical implications. 

The equation of state \eqref{eq_state} has a transparent physical interpretation. The first term, $s_0(\eta)$, has the meaning of the speed of a ``free'' (isolated) breather or soliton with the spectral parameter $\eta$. Indeed, in Eq.~\eqref{eq_state_sol} $s_0(\eta)=-4 \Re \eta$, which coincides with the group velocity $c_g$ of the fNLS fundamental soliton \eqref{nls_soliton} while in 
Eq.~\eqref{eq_state_breather} $s_0(\eta)=-2  \frac{\Im[\eta R_0(\eta)]}{\Im R_0(\eta)}$   is the group velocity  of the TW breather \eqref{TW_speed} (one sets $\delta_0 = iq$ in Eq.~\eqref{eq_state_breather}). The second (integral) term in \eqref{eq_state} describes the modification of the  ``tracer'' breather (soliton) mean velocity  due to its interaction  with other breathers (solitons) in the gas.  The interaction kernel $\Delta(\eta, \mu)$ for the soliton gas (cf. the second expression in Eq.~\eqref{eq_state_sol}) coincides with the   well-known expression for the position shift in the two-soliton interactions \cite{Zakharov:72, GEl:05}.  We  
{then conclude} that  Eq.~\eqref{eq_state_breather}
describes the position shift in a two-breather interaction. {The expressions for the position shifts in the interaction of two TW breathers were recently obtained in Refs. \cite{gelash_formation_2018}, \cite{li_soliton_2018} in a different, less explicit, form.  While we could not see an obvious way to verify the equivalence between the two representations analytically,  we have undertaken a numerical comparison   with the representation obtained in \cite{gelash_formation_2018} for a range of parameters, which convincingly confirmed  full agreement between the two. } 

{The fact that  the pairwise interaction kernels \eqref{eq_state_breather}, \eqref{eq_state_sol} show up in the equations of state  \eqref{eq_state}  without assuming any dilute nature of the gas implies that properties of breather and soliton gases are fully determined by  the ``fundamental''  two-particle interactions for the whole range of admissible densities.  We also note that the spectral thermodynamic limit  only yields the soliton-soliton (breather-breather) interaction kernel but not the kernel related to the interaction of solitons (breathers) with the continuous spectrum component, confirming thus the original premise of our paper that the thermodynamic limit of finite-gap potentials corresponds to  a ``genuine'' soliton (breather) gas.}

Finally, we  note that 
the  inequality $\sigma(\eta)  \ge 0$ in Eq.~\eqref{dr_breather_gas1} imposes a fundamental constraint 
\begin{equation}\label{const_u}
 \int_{\G^+} \Delta(\eta, \mu)
u(\mu) |d \mu| \leq 1
\end{equation}
on the function $u(\eta)$.

\medskip
 {\it Example:  Multi-component breather gas.} 
 
  Consider a breather gas  characterized by the density of states in the form of a linear combination of Dirac delta-functions centered at different spectral points $\eta^{(j)}$ (we use an upper index to distinguish these spectral points from the centers of spectral bands $\eta_j$, used earlier in the thermodynamic scaling construction)
 \be\label{u_delta1}
 u(\eta) = \sum \limits_{j=1}^{M} w_j \d(\eta - \eta^{(j)}),
 \ee
 where $w_j>0$ are the given components' weights and $\Im \eta^{(j )}> 0 \ \forall j$.  The structure of the multi-component reduction of the  equation of state for generalized soliton gas of the KdV type (i.e. when the integration in \eqref{eq_state} occurs along interval of the real line) has been  studied in \cite{el_kinetic_2011}. For a particular case of a two-component ($M=2$) fNLS soliton gas such a reduction has been considered in \cite{GEl:05}. Here we present a straightforward extension of the results of \cite{el_kinetic_2011}, \cite{GEl:05} to breather gas.
 
 Substituting  \eqref{u_delta1} into the equation of state \eqref{eq_state} we obtain a linear system for the gas' component velocities $s^{(j)}\equiv s(\eta^{(j)})$,  
 \begin{equation}\label{s_alg}
 s^{(j)}=s_0^{(j)} + \sum \limits_{m=1, m \ne j}^M \Delta_{jm} w_m(s^{(j)}- s^{(m)}), \quad j=1, 2, \dots M,
 \end{equation}
 where $s_0^j\equiv s_0(\eta^{(j)})=s_{\rm TW} (\eta^{(j)})$,  $\Delta_{jm}=\Delta(\eta^{(j)}, \eta^{(m)})$ (cf.~\eqref{eq_state_breather} for $\Delta(\eta, \mu)$
 in breather gas).
 For $M=2$ system \eqref{s_alg} can be solved explicitly to give
  \be \label{s_12}
  \begin{split}
 s^{(1)}= s_0^{(1)} + \frac{ \Delta_{12} w_2 (s_0^{(1)}-s_0^{(2)})}{1-(\Delta_{12} w_2+ \Delta_{21} w_1)}, \\
 \quad s^{(2)}= s_0^{(2)}-\frac{ \Delta_{21} w_1 (s_0^{(1)} - s_0^{(2)})}{1-(\Delta_{12} w_2+ \Delta_{21} w_1)}. 
 \end{split}\ee
An important remark is in order on the meaning of the delta-function ansatz \eqref{u_delta1}  for the density of states $u(\eta)$.  As a matter of fact, the representation \eqref{u_delta1}  is a mathematical idealisation, which has a formal sense in the context of the integral equation of state \eqref{eq_state},  but cannot be applied to the original dispersion relations 
 \eqref{dr_breather_gas1}, \eqref{dr_breather_gas2}  where it appears  in both the integral and the secular terms.  In a physically realistic description  the delta-functions in \eqref{u_delta1} should be replaced by some narrow distributions around the spectral points $\eta^{(j)}$, {i.e. we first take
the limit $N\to\infty$ and then later allow the distributions to become sharply peaked.}  As a result equation \eqref{dr_breather_gas1} would impose
 a constraint    \eqref{const_u} on $u(\eta)$ which, among other things,  implies  that the denominators in \eqref{s_12} must be positive.     Other constraints could arise due to the requirements of non-negativity for some statistical parameters of the  fNLS field (see \cite{el_critical_2016} for the  KdV soliton gas consideration), but we do not consider them here.

\section{Propagation of an isolated soliton/breather through a  gas}
The equation of state \eqref{eq_state} can be used to describe the propagation of an isolated soliton (or breather) with the spectral parameter $\eta \notin \G^+$ through a soliton (breather)  gas with  known density of states $u(\mu)$ (and the corresponding velocity $s(\mu)$ by \eqref{eq_state}), where $\mu \in \G^+$. 
We shall call such an isolated soliton (breather) a {\it trial} soliton (breather) (not to confuse with the {\it tracer} soliton (breather) with $\eta \in \G^+$). Apart from being a convenient tool for the numerical verification of the developed theory (see the relevant comparisons \cite{carbone_macroscopic_2016} for the KdV soliton gas case), the known dependence of the trial soliton (breather) velocity on $\eta$ (obtained, e.g., from a series of measurements of the mean velocity of trial  solitons (breathers) of different amplitudes propagating through the same gas) can be used for posing the inverse problem: recover the density of states 
$u(\mu),~\mu\in\G^+$ of a breather/soliton gas  from a given function $s(\eta),~ \eta\not\in\G^+$, i.e., determine  the gas by irradiating
it with the trial breathers/solitons and measuring  velocities of their propagation through the gas.   

Expressing $s(\eta)$ from \eqref{eq_state} we obtain the expression for the mean velocity of such a trial breather/soliton propagating through a respective gas:
\begin{equation}\label{speed_trial}
s(\eta)=\frac{s_0(\eta) - \int_{\G^+}\Delta(\eta, \mu)  u(\mu) s(\mu) |d \mu|}{1- \int_{\G^+}\Delta(\eta, \mu)  u(\mu) |d \mu|}.
\end{equation}
We note that technically,  equation  \eqref{eq_state} is valid only for $\eta\in\G^+$,
but we can always assume that an isolated point $\eta\in\C^+$ can be added to $\G^+$ with $u(\m)=w\d(\m-\eta)$ near that point.  Then, substituting this ``extended'' $u(\mu)$ in \eqref{eq_state} and taking the limit $w\ra 0^+$ 
 we obtain \eqref{speed_trial}. 
 
 We would like to stress an important subtlety associated with Eq.~\eqref{speed_trial} that can be easily overlooked. For $\eta \in \G^+$ Eq.~\eqref{speed_trial} is equivalent to the equation of state \eqref{eq_state} and thus represents an integral equation for $s(\eta)$. However, for $\eta \notin \G^+$, the right-hand side of \eqref{speed_trial} is assumed to be already known (from the solution of Eq.~\eqref{eq_state}) so  in that case Eq.~\eqref{speed_trial} represents an expression for $s(\eta)$ rather than an equation to be solved. 
 
 \bigskip
{\it Example:  \ Propagation of a breather through a one-component breather gas.}

 \medskip
Consider a trial breather with the spectral parameter $\eta=\eta^{(1)}$, where $\Im \eta^{(1)} >  |\delta_0|$, propagating through a one-component breather gas with $u(\eta)= w_2 \delta(\eta-\eta^{(2)})$, $s(\eta^{(2)})  = s_0(\eta^{(2)}) \equiv s_0^{(2)}$ (it is clear that in a one-component gas due to the absence of interactions the velocity of the gas coincides with the free soliton velocity, see Eq.~\eqref{speed_trial} with the interaction terms removed).
From Eq.~\eqref{speed_trial} we obtain
 \be \label{eq:43}
 s^{(1)}= \frac{s_0^{(1)}-\Delta_{12}w_2 s_0^{(2)}}{1-\Delta_{12}w_2},  
 \ee
 which is consistent with Eq.~\eqref{s_12} in the limit $w_1 \to 0^+$.

{\it Peregrine gas.} \  Consider  now the case when $\eta^{(2)}=\d_0=iq$,
where $q\in \R^+$, i.e., the ``mass'' $w$ of the delta function  is at the end of the exceptional band $\g_0=[-\d_0, \d_0]$. This is the ``spectral portrait'' of the Peregrine gas.  Then the logarithmic kernel in \eqref{dr_breather_gas1}, evaluated at $\m=\d_0$, is
\be \label{perergine_kernel}
-\ln \le|\frac{\d_0-\eta}{\d_0-\bar\eta}\ri|
+\ln \le|\frac{\eta\d_0-\d_0^2}
{\bar\eta\d_0-\d_0^2}\ri|=0,
\ee
where $\eta$ is any point in the upper halph-plane that has a branch cut $[0, \delta_0]$. Thus,  by \eqref{eq_state_breather},  $\Delta_{12}=0$ and equation \eqref{eq:43} yields $s^{(1)}(\eta) = s_0^{(1)}(\eta)$.
 i.e. the group velocity of the (free) TW breather. Thus, the propagation of the trial TW breather through a Peregrine gas is {\it ballistic}. 
 
 \medskip
 {\it Kuznetsov-Ma breather gas.} \ For the KM breather gas with $\eta^{(2)}=ip$, where $p>  q$, we have $s_0^{(2)}=0$ (see \eqref{TW_speed}), 
 so the propagation speed of a trial breather through a KM gas 
 is found from Eq.~\eqref{eq:43} to be $s^{(1)}= {s_0^{(1)}}/{(1-\Delta_{KM}w_2)}$, where 
 $ \Delta_{\mathrm KM}=\Delta_{12} $ with $\m=ip$ and $p>  |\delta_0|$. 
 
 \medskip
{\it Akhmediev breather gas.} \ Finally, for the AB gas with $\eta^{(2)}=ip \pm \epsilon$, where $p \in (0, |\delta_0|)$, $\epsilon \ra 0$, we have 
 $s_0^{(2)} \ra \pm \infty$  (see \eqref{TW_speed}).
 Denote the interaction kernel $\Delta_{12} \equiv \Delta_{\rm AB}$. 
 Then the velocity of a trial breather propagating through a AB gas is $s^{(1)} \sim -\frac{\Delta_{\rm AB}w_2 }{1-\Delta_{\rm AB}w_2} s_0^{(2)}$ as $\epsilon \to 0$. {Note that due to the infinite spatial extent of the Akhmediev breather one should require   that  the density of the AB gas $w_2 \to 0$ so that $w_2 s_0^{(2)}= \mathcal{O}(1)$ in the latter expression thus ensuring finite velocity of the test breather.}

\bigskip
The breather/soliton interactions in a breather/soliton gas not only modify the ``particle'' group velocity $c_g$ of a tracer breather (soliton)  but they also change the ``wave'' phase velocity of its carrier $c_p$. The phase velocity distribution in the carrier of a breather gas can be naturally defined  as $\tilde s (\eta)=\tilde v/ \tilde u$, where  $\tilde u(\eta)$ and $\tilde v(\eta)$ are the continuous carrier wavenumber and frequency functions respectively,  satisfying the  dispersion relations  \eqref{dr_breather_gas1p} (breather gas) and \eqref{dr_soliton_gas1p} (soliton gas).
As a result, we obtain
\begin{equation}\label{phase_velocity_BG}
\tilde s(\eta) = \frac{\tilde s_0 - \int_{\G^+} \tilde \Delta(\eta, \mu) u(\mu) s(\mu) |d\mu|}{1- \int_{\G^+} \tilde \Delta(\eta, \mu) u(\mu)  |d\mu|},
\end{equation}
where for breather gas: 
\bea
&&\tilde s_0=-\frac{2 \Re [\eta R_0(\eta)]}{\Re [R_0(\eta)]}, \label{eq_state_breatherp}  \\
&& \tilde\Delta(\eta, \mu) =\frac{1}{\Re [R_0(\eta)]} \le[\arg \frac{\eta-\m}{\eta-\bar\m} \right. \nonumber\\   
&& \left. -\arg\frac{R_0(\eta)R_0(\m)+\eta \m -\d_0^2}
{R_0(\eta)R_0(\bar\m)+\eta \bar\m-\d_0^2}\ri], \nonumber
\eea
and for soliton gas:
\begin{equation}\label{eq_state_solp}
\tilde s_0=-\frac{2 \Re [\eta^2]}{\Re \eta}, \quad \tilde \Delta(\eta, \mu) =\frac{1}{\Re \eta}\le[\arg \frac{\eta-\m}{\eta-\bar\m} - 2 \arg \mu \ri].
\end{equation}
{ (see the comment before Eq.~\eqref{dr_breather_gas1p}  with regard to the choice of the branches of  the $\arg$ functions in \eqref{eq_state_breatherp}, \eqref{eq_state_solp})}.

The expressions  for $\tilde s_0(\eta)$ in \eqref{eq_state_breatherp} and \eqref{eq_state_solp} coincide with phase velocities $c_{p}$ of a carrier in a TW breather (cf. \eqref{TW_speed}) and fundamental soliton (cf. \eqref{nls_soliton}) respectively. The interaction kernels $\tilde \Delta$ are clearly related  to (but don't coincide with) the expressions of the phase shifts in two-breather (two-soliton) interactions (see \cite{Zakharov:72}, \cite{gelash_formation_2018}, \cite{li_soliton_2018}).

One cannot help noticing the  similarity between  Eqs.~ \eqref{speed_trial}  and \eqref{phase_velocity_BG} defining respectively the group and phase velocity of a ``trial'' or ``tracer'' breather (soliton) in a breather (soliton)  gas. We need to stress however, that, despite the apparent similarity, these expressions have very different structure. 
Indeed Eq.~\eqref{speed_trial} contains only one type of velocities (the group velocities) while Eq.~\eqref{phase_velocity_BG} connects two types of velocities.
Also Eq.~\eqref{speed_trial} has a different meaning depending on whether $\eta \in \G^+$ (a ``tracer'' breather/soliton)  or $\eta \notin \G^+$ (a ``trial'' breather/soliton) (see the discussion after Eq. \eqref{speed_trial}), while Eq.~\eqref{phase_velocity_BG} does not distinguish between these two types of propagating breathers/solitons.

\section{ Rarefied breather/soliton gas  and soliton condensate}
The nonlinear dispersion relations \eqref{dr_breather_gas1} 
and \eqref{dr_soliton_gas1} for breather and soliton gas respectively were derived assuming the general, exponential spectral scaling \eqref{exp_scal} implying that generically,  
the integral and the secular terms in these relations are of the same order. In the other two scalings considered in Section \ref{sec-therm} one of these two terms must be subdominant:
this will be the integral term in the super-exponential scaling and the non-integral term in the sub-exponential scaling. 

It is convenient to characterize the spectral scalings and the corresponding breather/soliton gases in terms of the function $\sigma(\eta)$ parametrizing the dispersion relations \eqref{dr_breather_gas1} and \eqref{dr_soliton_gas1}. From Eq.~\eqref{dr_breather_gas1} we have for breather gas:
\begin{equation}\label{sigma_bg}
 \sigma (\eta)=\frac{\Im [R_0(\eta)]( 1-\int_{\G^+} \Delta(\eta, \mu)
u(\mu) |d\mu|)}{u(\eta)} \ge 0,
\end{equation}
where the interaction kernel $\Delta(\eta, \mu)$ is given by Eq.~\eqref{eq_state_breather}. The expression for $\sigma(\eta)$ in a soliton gas is obtained from Eq.~\eqref{sigma_bg} by replacing $\Im R_0(\eta)$ with $\Im \eta$ and using Eq.~\eqref{eq_state_sol} for 
$\Delta(\eta, \mu)$.  For the exponential scaling $\sigma(\eta)=\mathcal{O}(1)$, while the limiting cases $\sigma \to \infty$ and $\sigma \to 0$ correspond to the super- and sub-exponential spectral scalings respectively.

 \subsection{Rarefied breather/soliton gas}\label{sec-rarefied}

Rarefied breather/soliton gas represents an infinite random ensemble of  weakly interacting breathers/solitons characterized by small density of states, $u \ll 1$, and therefore, $\sigma \gg 1$ by \eqref{sigma_bg}.  We shall refer to the limit $u \to 0$,  $\sigma \to \infty$,  $u \sigma = \mathcal{O}(1)$ as the {\it ideal gas limit} as it corresponds to the gas of non-interacting breathers (solitons). Spectrally this limit corresponds to the super-exponential spectral scaling \eqref{sup-exp_scal}.  For a rarefied gas the interaction (integral) term in the equation of state \eqref{eq_state} is sub-dominant so the leading order term $s(\eta) = s_0(\eta)$ describes the group velocity distribution in an ideal breather (soliton) gas. Then the  first correction to the ideal gas velocity $s_0(\eta)$ is readily computed to give:
\begin{equation}\label{rarefied_group}
s(\eta) \approx  s_0(\eta) + \int \limits_{\G^+}\Delta(\eta, \mu)[s_0(\eta) - s_0(\mu)] u(\mu) |d \mu|.
\end{equation}
Eq.~\eqref{rarefied_group} represents a fNLS breather/soliton gas counterpart of the equation for the soliton velocity in a rarefied KdV soliton gas obtained in \cite{Zakharov:71}.
Similarly,  the carrier phase velocities  in the ideal gas are determined by the leading order term $\tilde s(\eta) = \tilde s_0(\eta)$ in \eqref{phase_velocity_BG}, while the correction due to weak interactions in a rarefied gas yields 
\begin{equation}\label{rarefied_phase}
\tilde s (\eta) \approx  \tilde s_0(\eta) + \int \limits_{\G^+} \tilde \Delta(\eta, \mu)[\tilde s_0(\eta) -  s_0(\mu)] u(\mu) |d \mu|.
\end{equation}
(Note that, unlike in \eqref{rarefied_group}, the integral term in \eqref{rarefied_phase} involves the difference between the phase and group velocities of a free breather/soliton).

In the ideal gas limit, the interaction terms in the complex nonlinear dispersion relations \eqref{dr_breather_gen1} for breather gas are subdominant  leading to
\be\label{idgas-main}
\tilde u(\eta)-i\sigma(\eta)u(\eta) = -R_0(\eta),\quad  \tilde v(\eta)-i\sigma(\eta)v(\eta) = 2\eta R_0(\eta).
\ee
Taking the real and imaginary parts of \eqref{idgas-main} we recover the expressions \eqref{eq_state_breather} and \eqref{eq_state_sol} for the group velocity $s_0(\eta)$ and \eqref{eq_state_breatherp}, \eqref{eq_state_solp} for the carrier phase velocity $\tilde s_0$  in the ideal breather gas (soliton gas in the limit $\d_0 \to 0$) as expected. We also observe that the ratio
 \begin{equation}\label{crossover}
\frac{\tilde v(\eta)-i\sigma(\eta)v(\eta)}{\tilde u(\eta)-i\sigma(\eta)u(\eta)}= -2\eta
\end{equation}
is the same for  breather and soliton gas in this regime.

We note that the discrete ($2N$-gap solution) counterpart of \eqref{crossover} 
\be\label{rem-rel}
\frac{\tilde \o_j+\frac{2i\ln |\d_j|}{\pi }\o_j}{\tilde k_j+\frac{2i\ln |\d_j|}{\pi }k_j}=-2\eta_j, \quad j=1,\dots, N
\ee
suggests the following  wavenumber-frequency scaling: 
\be \label{crossover1}
k_j \sim \o_j \sim\ln^{-1}|\delta_j|,
\ee
that  is generally different from the one given by Eq.~\eqref{scal_k_nu} (1D) and \eqref{scal_k_nu2} (2D), as it does not involve $N$. In particular, the  wavenumber-frequency scaling \eqref{crossover1}, together with super-exponential spectral scaling \eqref{sup-exp_scal}  covers the case of the transition from the $2N$-gap solution to the $N$-soliton solution for $N$ fixed.

  \subsection{Soliton/breather condensate} \label{sec-sol_cond}
    
It has been already  mentioned that the  inequality $\sigma(\eta)  \ge 0$ in Eq.~\eqref{sigma_bg} imposes a fundamental constraint 
on the density of states  $u(\eta)$. As discussed in Sec.~\ref{therm_1D}, the critical value $\sigma(\eta)=0$  corresponds to the sub-exponential spectral scaling \eqref{sub-exp_scal}.  One can see from the nonlinear dispersion relations \eqref{dr_breather_gas1}, \eqref{dr_breather_gas2}  that in this case the gas properties are fully determined by the interaction (integral) terms, while the information about the individual quasi-particles  (described by the secular terms) is completely lost.  By analogy with Bose-Einstein condensation we shall call the breather (soliton) gas at $\sigma=0$ the {\it breather (soliton) condensate}. From \eqref{sigma_bg}  we obtain the criticality condition 
 \begin{equation}\label{crit_u}
 \int_{\G^+} \Delta(\eta, \mu)
u(\mu) |d \mu| =1,
\end{equation}
which is the limiting case of the constraint \eqref{const_u}.
For a given interaction kernel $\Delta(\eta, \mu)$ (Eq.~\eqref{eq_state_breather} for breather gas and Eq.~\eqref{eq_state_sol} for soliton gas)  Eq.~\eqref{crit_u} represents an integral equation (the Fredholm integral equation of the first kind) for the critical density of states $u=u_{\rm c}(\eta)$.  

{Equation \eqref{crit_u} admits a lucid physical interpretation. Indeed, introducing the total density of states  in the condensate, $\rho_c = \int_{\G^+}  u_c(\mu) |d \mu|$, we re-write \eqref{crit_u} as
\begin{equation}\label{cond_phys}
\langle \Delta \rangle =\rho_c^{-1},
\end{equation}
where $\langle \dots \rangle$ denotes averaging over the spectral measure $\rho_c^{-1} u_c(\eta)$. Equation \eqref{cond_phys}  then implies that in a soliton (breather) condensate the average position shift due to collisions is equal to the average distance between quasiparticles.
}

If the spectral locus curve $\G$ belongs to a vertical line (a bound state gas), equation  \eqref{crit_u} with the soliton gas interaction kernel \eqref{eq_state_sol} can be solved explicitly  using the inversion formula for the Hilbert transform. Another explicit solution for the   density of states in soliton condensate can  be obtained for the special case when the curve $\G$ represents a circle or a circular arc in the complex plane. Below we consider these two important examples.

 \medskip

 {\it Example 1. Bound state soliton condensate} 
 
Let $\delta_0=0$ (soliton gas case) and  $\G=[-iq,iq]$ for some $q>0$, which we shall call the soliton condensate intensity.
 Then $v(\eta)\equiv 0$ solves \eqref{dr_soliton_gas2}. The remaining equation \eqref{dr_soliton_gas1}  with $\s\equiv 0$  and $\eta \in \G$ can be re-written as
 \be \label{eq:54}
 \int_{-iq}^{iq}\ln|\m-\eta|u(\m)id\m = \Im \eta,
 \ee
 where  we assume an odd (anti Schwarz symmetric) extension of $u(\m)$ onto $[-iq, 0]$. {Indeed, 
 \be
 \ln \le|\frac{\m-\bar\eta}{\m-\eta}\ri|u(\m)= - \le[\ln|\m-\eta|u(\m) +  \ln|\bar\m-\eta|u(\bar\m)\ri], 
 \ee
 where we have assumed that $u(\bar\m)=-u(\m)$ for all $\m\in[0,iq]$.}
 Introducing new variables $\x=\Im \eta$, $y=\Im \m$
 we obtain
 \be \label{int_H}
  -\int_{-q}^{q}\ln|\x-y|\hat u(y)dy = \x,
 \ee
 where $\hat u(y) = u(iy)$. Differentiating \eqref{int_H}, we obtain 
\be
\pi H[\hat u](\x):=\int_{-q}^{q}\frac{\hat u(y)dy}{y-\x} = 1,
 \ee
 where $ H[\hat u]$ denotes the finite Hilbert transform (FHT) of $\hat u$ over $[-q,q]$  \cite{tricomi_book}. Inverting the FHT $H$ subject to an additional constraint
 $H[\hat u](0)=0$ (see, e.g., \cite{okada_finite_1991}), we obtain  the density of states for the bound state soliton condensate
 \be\label{Weyl}
 u(\eta)=\frac {-i \eta }{ \pi\sqrt {\eta^2+q^2}} , ~~~\eta \in (-iq,iq).
  \ee
 One can observe that the density of states \eqref{Weyl} of the bound state soliton condensate   coincides with the appropriately normalized semi-classical distribution of discrete ZS spectrum for a rectangular barrier,  obtained as a derivative of the corresponding Weyl's law following from the Bohr-Sommerfeld quantisation rule  for the ZS operator (see \cite{Zakharov:72}, \cite{Jenkins:14} and references therein). 

 We now use Eqs.~\eqref{dr_soliton_gas1p} to evaluate the bound state soliton condensate carrier wave parameters $\tilde u,\tilde v$.
Since $\arg \m=\frac \pi 2$ and 
\be
\arg \frac{\eta-\m}{\eta-\bar\m}=-\pi \chi_{[\eta,iq]},
\ee
where $\chi$ denotes the characteristic function and $\m,\eta\in[0,iq]$, we calculate 
\be \label{eq:59}
\tilde u(\eta)= -(\Re \sqrt{q^2+\eta^2}+q),~~~~\tilde v(\eta)= 2 \Re(\eta^2),
\ee
where we used the fact that the integral term in \eqref{dr_soliton_gas1p} is zero due to $v(\eta)=0$ when $\eta \in [-iq, iq]$. The carrier phase velocity in the bound state soliton condensate is then
\be \label{phase_cond}
\tilde s(\eta) = \frac{\tilde v(\eta)}{\tilde u(\eta)}=
-\frac{2 \Re(\eta^2)}{\Re \sqrt{q^2+\eta^2}+q}.
\ee
As a matter of fact, Eq.~\eqref{phase_cond} could be obtained directly from Eq.~\eqref{phase_velocity_BG}. When $q \to 0$ we recover from \eqref{phase_cond} the carrier phase velocity of the fundamental soliton \eqref{nls_soliton} (which should be treated as a ``trial'' soliton here).

To evaluate the speed of a trial soliton with $\eta \notin [-iq, iq]$  propagating through the bound state soliton condensate we substitute \eqref{Weyl} in \eqref{speed_trial} and arrive at 
\be\label{s-vert-in}
s(\eta)= \frac{ -4\Im \eta\Re \eta}{\Im\eta-\frac {1 }{\pi} \Re\int_{-iq}^{iq}\ln (\m-\eta)\frac { \m d\m }
{\sqrt {\m^2+q^2}}}.
\ee
 Applying integration by parts and evaluating residues, we obtain
 \be\label{s-vert0}
s(\eta)=- \frac{ 4\Im \eta\Re \eta}{\Im
\sqrt {\eta^2+q^2}}.
\ee
It is interesting to compare the group velocity  \eqref{s-vert0} of the trial soliton propagating though a soliton condensate of intensity $q$  with  the  group velocity  $c_g$ of the TW breather with the same soliton eigenvalue $\eta$ and the same background intensity $q$ (see Eq.~\eqref{TW_speed}). One can see that, although the expressions for these  two velocities are different,  they exhibit the same asymptotic behavior up to the second order terms: $s(\eta)=-4\Re \eta (1-d^2/2) + \mathcal{O}(d^4) $, where $d= q/\Im \eta \ll 1$ and $\Re \eta = \mathcal{O}(1)$.

 Concluding this important example, we note that  the very recent numerical study \cite{gelash_bound_2019} has shown that statistical characteristics (the probability density function, the Fourier power spectrum and the autocorrellation function) of the bound state soliton gas modelled by  an $n$-soliton solution with $n \gg 1$ and the ``Weyl'' density of states  \eqref{Weyl} agree with remarkable accuracy with the counterpart characteristics of the stationary integrable turbulence describing the asymptotic, long-time behaviour of the spontaneous modulational instability (i.e. the modulational instability of a plane wave of intensity $q$ perturbed by a small noise) {studied numerically  \cite{Agafontsev:15}) and experimentally \cite{kraych_statistical_2019}}.

 \bigskip

  {\it Example 2. \ Non bound state ``circular''  soliton condensate} 
  
We now present an example of soliton condensate which is not a bound state, i.e. a dynamic soliton condensate.   
Consider the spectral locus curve $\G$ in the form of an arc of the circle $|\eta|=\r>0$ connecting, counterclockwise, the points $\bar \eta_1$  and $\eta_1$ of $\G$, where $|\eta_1|=\r$. 
  
To solve Eq.~\eqref{crit_u} in this case we introduce the 
   change of  
variables $\mu=\r e^{i\th}$,
$\eta=\r e^{i\x}$.
Then  \eqref{crit_u}     becomes    
\be
\r\int_{0}^{\x_1}\ln\le|\frac{\sin \frac{\x-\th}{2}}{\sin \frac{\x+\th}{2}}\ri|\hat u(\th)d\th=-\r\sin\x,
\ee
where $\hat u(\theta) = u(\r e^{i \th})$,  $\eta_1=\r e^{i\x_1}$. We  now   differentiate both sides in $\xi$ to obtain
\be\label{superdense_x-eqv}
\frac{1}{\pi}\int_{b}^{1}\frac{f(q)}{q-p}dq =-\frac p\pi,
\ee
where we introduced the new variables $p=\cos\xi$, $q=\cos\th$ and notations
$b=\cos\x_1$ and $f(p)= \hat u(\cos^{-1}p)$.  The integral in the left hand part of equation \eqref{superdense_x-eqv} represents the FHT $H[f](p)$ of the function $f$ over $[b,1]$, see
\cite{tricomi_book}, \cite{okada_finite_1991}. To ensure  uniqueness of the FHT inversion in \eqref{superdense_x-eqv} we impose a constraint $f(1)=0$, which is equivalent to a physically natural condition $\hat u(0)=0$.
Omiting the calculations, we present the result:
\bea \label{eq:56}
&& f(p)=- \frac{(p-1)(p+\frac{1-b}2)}{\pi R_+(p)},  \\
 &&  R_+(p)=\sqrt{(1-p)(p-b)}, \nonumber
\eea
assuming the positive value of the radical.
Note that $f(p)>0$ for $p\in[b,1)$.  Thus, we obtain the density of states in the circular soliton condensate
\bea
&& u(\eta)=\frac 1\pi\sqrt{\frac{1-\cos\x}{\cos\x-\cos\x_1}}(\cos\x+\frac{1-b}2) \nonumber \\
&&=
\frac 1\pi\le(\frac{\Re\eta}{|\eta|}+\frac{1-b}2\ri)
\sqrt{\frac{|\eta|-\Re\eta}{\Re\eta-\Re\eta_1}}. \label{superdense_x}
\eea

Equation \eqref{dr_soliton_gas2} with $\sigma=0$ for $v(\eta)$ for this condensate is solved in a similar fashion. As a result,  the soliton group velocity $s(\eta)=v(\eta)/u(\eta)$ is obtained  in the form:
\bea
&&s(\eta)=-8\r\frac{\cos^{(2)}\x+\frac{1-b}2\cos\x-\frac{(b+1)^2}8}{\cos\x+\frac{1-b}2}  \nonumber \\
=
&&-\frac{8{(\Re\eta)^2}+4(1-b){|\eta|}\Re\eta - (b+1)^2|\eta|^2}{{\Re\eta}+\frac{1-b}2{|\eta|}}. \label{velocity-circle}
\eea
In particular, for $\G$ being a circle, i.e. $b=-1$, we have
\be \label{pcases}
u(\eta)=\frac{\Im \eta}{\pi \rho}, \quad  v(\eta)=-8\frac{\Re\eta\Im \eta}{\pi \rho},    \quad s(\eta)=-8\Re\eta.
\ee
The latter expression means that the soliton speed within the ``circular'' soliton condensate is exactly twice the speed $s_0(\eta)= - 4 \Re \eta$ of the free soliton with the same spectral parameter $\eta$. The opposite limit $b\ra 1^-$ (a small circular arc near the origin with $\cos\x\ra 1^-$)  yields $s(\eta) \approx -8\r\frac{\cos^2\x-\hf}{\cos\x}\approx -4\Re\eta$, which is consistent with the speed of free (non-interacting) 
fundamental soliton \eqref{nls_soliton}. 

Using Eqs.~\eqref{superdense_x}, \eqref{velocity-circle} for $u(\eta)$, $v(\eta)=u(\eta)s(\eta)$  the wavenumber $\tilde u(\eta)$ and frequency $\tilde v(\eta)$  of the carrier wave in the ``circular'' condensate can be evaluated from Eqs.~\eqref{dr_soliton_gas1p}.

 \subsection{From ideal soliton gas to soliton condensate}

The explicit results for soliton condensates obtained in the previous subsection can be readily generalized to some ``genuine''  (non-condensate) soliton  gases, enabling one to span a continuum of states of  increasing density between an ideal soliton gas and a soliton condensate. We shall obtain the corresponding results for the bound state soliton gas  and the dynamic circular soliton gas following Examples 1 and 2 above.

\medskip
 We use the bound state soliton condensate solution \eqref{Weyl}    of the integral equation  \eqref{eq:54} (Eq.~\eqref{dr_soliton_gas1} with $\sigma \equiv 0$)  to derive a particular solution to the full Eq.~\eqref{dr_soliton_gas1} with $\s \ne 0$.  Let  $u=u_{\rm c}(\eta)$ be  the ``Weyl'' distribution \eqref{Weyl}  and assume that $\s(\eta) u_{\rm c}(\eta) = m\Im\eta$
 for some $m \geq 0$. Then $u(\eta)=u_{\rm c}(\eta) /(m+1)$ is the solution of Eq.~\eqref{dr_soliton_gas1} with 
  $\s(\eta)=m\pi\sqrt {\eta^2+q^2} $.
One can  see that  in the limit $m\ra 0^+$ one has $u \to u_{\rm c}$, i.e. the soliton gas approaches the state of condensate (cf. Eq.~\eqref{Weyl}), whereas 
in the limit $m\ra+\infty$ it approaches the state of ideal gas with $u(\eta) \to 0^+ $, $ \s(\eta)\ra +\infty$,  $u \s \to \pi u_{\rm c}\sqrt {\eta^2+q^2}=-i\eta$.

We now  derive the speed of a trial soliton with spectral parameter $\eta$, moving through the bound state soliton gas on $[-iq,iq]$.  We substitute the density of states $u=u_{\rm c}/(m+1)$  in the trial soliton speed formula Eq.~ 
 \eqref{speed_trial} to obtain (cf. Eqs.~\eqref{s-vert-in}, \eqref{s-vert0})
\bea
&&s(\eta)= \frac{ -4\Im \eta\Re \eta}{\Im\eta-\frac {1 }{\pi(m+1)} \Re\int_{-iq}^{iq}\ln (\m-\eta)\frac { \m d\m }{\sqrt {\m^2+q^2}}}  \nonumber\\
&&=  \frac{ -4\Im \eta\Re \eta}{\frac{m}{m+1}\Im\eta-\frac {1 }{m+1} \Im
\sqrt {\eta^2+q^2}}, \label{s-vert-in1}
\eea
where, as earlier, we have used integration by parts and residues to obtain the final expression.

In particular, for a ``trial'' soliton propagating through a bound state soliton condensate ($m=0$) we recover Eq. \eqref{s-vert0}.  In the opposite limit $m \to + \infty$ (trial soliton propagating through an ideal soliton gas) we recover the free soliton speed $- 4 \Re \eta$.

Similar to the  condensate case, we use \eqref{dr_breather_gas1p}-\eqref{dr_breather_gas2p} to evaluate $\tilde u,\tilde v$ to obtain (cf. \eqref{eq:59})
\be
\tilde u(\eta)= -\frac{\Re \sqrt{q^2+\eta^2}+q}{m+1},~~~~\tilde v(\eta)= 2 \Re (\eta^2).
\ee
When $m \to \infty$ (ideal gas) we have $\tilde u =0$, $\tilde v(\eta)= 2 \Re (\eta^2)$ in full agreement with the parameters of an isolated soliton \eqref{nls_soliton}.
 
  \medskip
  The extension of the circular condensate results to the full range of regimes proceeds in the similar way. For simplicity we only present  the results for the circular soliton gas when the spectral locus curve $\G$ is  a complete circle $|\eta|=\rho$. To study the range of regimes  between an ideal soliton gas and the corresponding  soliton condensate, we  take the ``seed'' condensate solution $u_{\rm c} = \Im \eta / \pi \rho$ (Eq.~\eqref{superdense_x} with $b=-1$) for the density of states and 
choose $\s$ in the integral equation  Eq.~\eqref{dr_soliton_gas1} so that
$\s(\eta)u_{\rm c}(\eta)= m\Im \eta$,  where $m \geq 0$. Now, invoking the same arguments that we used for the description of the 
bound state soliton gas we obtain the density of states and the group velocity in the circular gas as 
 \bea
 && u(\eta)=\frac{\Im \eta}{\pi \r (m+1)},  \qquad v(\eta)=-8\frac{\Re\eta\Im \eta}{\pi \rho(2m+1)},   \nonumber \\   && s(\eta)=-8\Re\eta\cdot \frac{m+1}{2m+1}. \label{circle-uv}
   \eea

As expected, Eq. \eqref{circle-uv} shows the growing density of states $u$ for the circular soliton gas as $m$ varies from $m\ra +\infty$ (ideal gas) to $m=0$ (condensate).  In this range of $m$, the speed $s(\eta)$ of the tracer 
soliton interpolates between the speed of a free soliton $-4 \Re \eta$ at $m=\infty$ and the condensate speed \eqref{pcases} at $m=0$.

Let us now use \eqref{dr_breather_gas1p} to calculate $\tilde u,\tilde v$ for the circular gas.
Using elementary geometry, we obtain 
\be
\arg \frac{\eta-\m}{\eta-\bar\m}
  = -\arg \eta+ \pi \chi_\eta, 
\ee
  where $\chi_\eta$ denotes the characteristic function of the arc $(\arg\eta,\pi)$ of our circle. Then
 \be\label{int-for-u}
 \int \limits_{\r}^{-\r}\le[\arg \frac{\m-\eta}{\m-\bar\eta}
-2 \arg \m \ri] u(\m)|d\m|=\frac {\frac{2\r\arg \eta }{\pi}+\Re \eta -3\r}{m+1},
\ee
so that 
\bea\label{circ-tilde-uv}
\tilde u(\eta)&=&-\frac{m}{m+1}\Re \eta- \frac{\r}{m+1}\le[\frac{2\arg \eta }{\pi} +1\ri],\cr
\tilde v(\eta)&=&\frac{2(m-1)}{(m+1)}(\Re \eta)^2- 2 (\Im \eta)^2
\eea
where the integral for $\tilde v$ was calculated similarly to \eqref{int-for-u}. One can now obtain the phase speed
\be\label{circ-tilde-s}
\tilde s(\eta)=2\frac {(m+1)(\Im \eta)^2-(m-1)(\Re\eta)^2}{m\Re \eta +\r\le(\frac{2\arg \eta }{\pi} +1\ri)},
\ee
which exhibits the expected limit $-2\frac{\Re\eta^2}{\Re\eta}$ as $m\ra\infty$, compare with the phase speed of
free soliton \eqref{nls_soliton}. In the condensate regime, \eqref{circ-tilde-s}  yields
\be
\tilde s(\eta)=\frac{2\pi\r}{\pi + 2\arg \eta}.
\ee

\section{Kinetic equations for breather and soliton gas}

\subsection{General construction}
So far we have assumed that the spectral characteristics $u(\eta)$, $v(\eta)$, $\tilde u(\eta)$, $\tilde v(\eta)$ of a breather/soliton gas  do not depend on $x,t$. We call such breather/soliton gases uniform, or homogeneous,  gases. For a non-homogeneous breather/soliton gas we introduce  $u\equiv u(\eta, x,t)$, $v \equiv v(\eta, x,t)$, $\tilde u \equiv \tilde u(\eta, x,t)$, $\tilde v \equiv \tilde v(\eta, x,t)$ assuming that variations of the gas' parameters occur on much larger spatiotemporal scales $\Delta x, \Delta t$ than the typical scales $\Delta x \sim \Delta t = \mathcal{O}(1)$  of the oscillations corresponding to individual solitons or breathers.

To derive evolution equations for a breather/soliton gas we go back to the original, discrete wavenumber and frequency components
$k_j({\bs \alpha})$, $\omega_j ({\bs \alpha})$, $\tilde k_j({\bs \alpha})$, $\tilde\omega_j({\bs \alpha})$ of the finite-gap potential, defined in terms of of the fixed branch points ${\bs \alpha}$ of the Riemann surface $\mathcal{R}$ of \eqref{rsurf}. Let us now consider a slowly modulated finite-gap potential, so that  ${\bs \alpha}={\bs \alpha} (x,t)$.
The equations describing the evolution of the ${\bs \alpha}={\bs \alpha} (x,t)$---the so-called Whitham modulation equations \cite{whitham}---for the fNLS equation have been derived in  \cite{pavlov_nonlinear_1987} for $n=1$ and in \cite{dafermos_geometry_1986} for an arbitrary genus $n$. The resulting system of $2n$ quasi-linear modulation equations for the branch points $\a_j(x,t)$ has an infinite number of conservation laws, that include a  finite subset of $n$ ``wave conservation'' laws \eqref{wcl0}, which we write in terms of the special wavenumber-frequency set satisfying the nonlinear dispersion relations \eqref{WUPjM},
\begin{eqnarray}
\partial_t k_j({\bs \a})&=& \partial_x \omega_j ({\bs \a}), \quad j=1, \dots, N \label{wc}, \\
\partial_t \tilde k_j ({\bs \a})&=& \partial_x \tilde \omega_j({\bs \a}), \quad j=1, \dots, N  \label{wc_tilde}.
\end{eqnarray} 
(as earlier, we assume even genus $n=2N$).  {\ We fix the exceptional band $\gamma_0$, which is consistent with the Whitham equations as the branch points $\alpha_j$ are the analogs of Riemann invariants  \cite{dafermos_geometry_1986}}.

We now apply the thermodynamic wavenumber-frequency scaling  \eqref{scal_k_nu} to the conservation equations \eqref{wc},\eqref{wc_tilde}.
Let $K_M=\sum \limits_{j=1}^{M} k_j$, $W_M=\sum \limits_{j=1}^{M} \o_j$, where $1 \leq M \leq N$. For convenience we shall be using 1D spectral configuration while the result will remain valid for 2D.
Invoking the 1D scaling \eqref{scal_k_nu} we obtain 
\begin{equation}
K_M=\sum _{j=1}^{M} \frac{\kappa(\eta_j)}{N} \to \int \limits_{\eta_\infty}^{\eta} \kappa(\mu) \varphi(\mu) |d\mu| \equiv \mathcal{K}(\eta),
\end{equation}
where we made a replacement $\eta_M \to \eta$ in the continuous limit.
Now, using \eqref{dens_states} we see that $\mathcal{K}'(\eta)=u(\eta)$, where the prime denotes differentiation along $\G^+$. Thus, $\mathcal{K}(\eta)$ has the meaning of the integrated density of states in a breather/soliton gas. Similarly, we have  
\begin{equation}
W_M=-\sum _{j=1}^{M} \frac{\nu(\eta_j)}{N} \to - \int \limits_{\eta_\infty}^{\eta} \nu(\mu) \varphi(\mu) |d\mu| \equiv \mathcal{V}(\eta),
\end{equation}
so that $\mathcal{V}'(\eta)= -v(\eta)$. Now, the thermodynamic limit of \eqref{wc} yields the continuity equation for the density of states $u(\eta, x, t)$
\begin{equation}\label{transport}
\partial_t u + \partial_x (us)=0,
\end{equation}
where $u=u(\eta, x,t), \ s=v(\eta, x,t)/u(\eta, x, t)$, and the dependence  $s[u]$ is given by the equation of state \eqref{eq_state}. One can see from \eqref{transport} that the breather/soliton group velocity $s(\eta,x,t)$
has the meaning of  the transport velocity of the corresponding gas. Equation \eqref{transport},  together with the equation of state \eqref{eq_state}  form the kinetic equation for a breather gas (using Eq.~\eqref{eq_state_breatherp})
or soliton gas (using Eq.~\eqref{eq_state_solp}).

Similarly, for the carrier wave components we have $\tilde k_j \to \tilde u (\eta, x, t)$, $\tilde \omega_j \to - \tilde v(\eta, x, t)= \tilde u(\eta, x, t) \tilde s (\eta, x, t)$ in the thermodynamic limit and Eqs.~\eqref{wc_tilde} transform into a single equation for the carrier wavenumber $\tilde u$:
\begin{equation}\label{transport_tilde}
\partial_t \tilde u + \partial_x(\tilde u \tilde s) =0,
\end{equation}
where the phase velocity $\tilde s$  also plays the role of the transport velocity for the carrier wavenumber, and the dependencies $\tilde u [u]$, $\tilde s[u]$ are given by \eqref{dr_breather_gas1p}, \eqref{phase_velocity_BG}. Together with the carrier wave nonlinear dispersion relations \eqref{dr_breather_gas1p}
equation \eqref{transport_tilde} yields a ``satellite'' kinetic equation for the carrier wave, reflecting the 
the dual, ``wave-particle'' nature of breathers and solitons.  We stress that the previous works on soliton gas kinetic equations (see e.g. \cite{GEl:05, el_kinetic_2011, doyon_soliton_2018} and references therein) have been concerned only with the ``particle'' equation \eqref{transport}.

{One may note that the application of the thermodynamic limit to the Whitham modulation equations  is in apparent conflict with the original premise of the Whitham theory \cite{whitham}, that is based on the scale separation, in which
spatiotemporal modulation scales are large compared to multiperiodic wavelengths and periods. We however note that the soliton limit of the Whitham equations, while being formally inconsistent with the original assumptions of the modulation theory, is well known to yield an accurate  description of the dynamics of individual solitons and soliton trains in a variety of problems, including the dispersive shock wave theory \cite{el_dispersive_2016} and the recently introduced theory of soliton-mean flow interactions \cite{maiden_solitonic_2017}, \cite{sprenger_hydrodynamic_2018} confirmed  by  both direct numerical simulations and physical experiments. In fact, Whitham  in his book showed how to describe solitary wave trains by an appropriate limit of the modulation system, see \cite{whitham}, Ch. 16.16.}

\subsection{Multi-component hydrodynamic reductions}
\label{sec:multi}
The kinetic equation \eqref{transport}, \eqref{eq_state} can be reduced to a system of quasilinear PDEs if one takes advantage of the multi-component delta-function ansatz \eqref{u_delta1} for the density of states $u$, where one now assumes that the ``densities'' $w_j$ and the speeds $s^{(j)}$ of the gas 
components are  slow functions of $x, t$. As we mentioned in Sec.~\ref{sec:eq_state}, there are certain constraints on the admissible values of $w_j$; we shall assume that these constraints are satisfied.

 The resulting system for $w_j(x,t)$, $s^{(j)}(x,t)$, $j=1, \dots, M$ has the form of a system of hydrodynamic conservation laws 
\be \label{cont}
(w_j)_t + (w_j s^{(j)})_x=0, \qquad j=1, \dots, M
\ee
with the closure conditions given by the multicomponent equation of state Eq.~\eqref{s_alg}.

The hydrodynamic type system \eqref{cont}, \eqref{s_alg} was extensively studied in \cite{el_kinetic_2011}, where it was shown that it represents a {\it hyperbolic integrable linearly degenerate system for any $M \in \mathbb{N}$}. Thus, all the previously obtained general mathematical results for finite-component KdV-type soliton gases \cite{el_kinetic_2011} can be readily extended  to the case of fNLS soliton and breather gases.
In particular, as was shown in \cite{GEl:05, el_kinetic_2011} for $M=2$ the  system \eqref{cont}, \eqref{s_alg} reduces to 
\be \label{chaplygin}
(s^{(1)})_t + s^{(2)} (s^{(1)})_x=0, \quad (s^{(2)})_t + s^{(1)} (s^{(2)})_x=0
\ee
with the relations between $s^{(1,2)}$ and $w^{1,2}$ are given by Eq.~\eqref{s_12}.

System \eqref{chaplygin} represents the diagonal form of the  so-called Chaplygin gas equations,  the system of isentropic gas dynamics with  the equation of state $p= - {A}/{\rho}$, where $p$ is a pressure, $\rho$ is the gas density and $A>0$ is a constant. It  occurs in certain theories of cosmology (see e.g. \cite{bento_generalized_2002}) and is also equivalent to the 1D Born-Infeld equation \cite{born_foundations_1934, whitham} arising in nonlinear electromagnetic field theory.

{We note that hyperbolicity of the hydrodynamic system  \eqref{cont}, \eqref{s_alg} might look  surprising in the context of the fNLS equation as the fNLS-Whitham  system  is known to be elliptic for a generic set of modulation parameters and for any genus \cite{dafermos_geometry_1986}. The apparent paradox is resolved by noticing that the fNLS-Whitham system  exhibits real eigenvalues (characteristic speeds) in the soliton limit. E.g. for the genus 1 case, the two pairs of complex conjugate eigenvalues of the modulation matrix degenerate into a single, quadruply degenerate real eigenvalue, corresponding to the velocity of the fundamental soliton, see e.g. \cite{el_dam_2016}. }

Due to availability of conservation laws \eqref{cont} one can solve a general Riemann problem for a multicomponent-component breather gas described by \eqref{cont}, \eqref{s_alg}. As is known, a weak solution to the Riemann problem for a system of hydrodynamic conservation laws \eqref{cont} generally consists of $M+1$ disparate constant states separated by $M$ propagating discontinuities  or rarefaction waves (one of each family), where the jumps of $w^i$ and  $s^i$ across the discontinuities are determined from the Rankine-Hugoniot conditions  \cite{lax_hyperbolic_1973}.  Linear degeneracy of the system \eqref{cont}, \eqref{s_alg}  implies that there are no rarefaction waves, and the shocks are contact discontinuities with the speeds coinciding with the speeds of the relevant components \cite{rozhdestvenskii_systems_1983}. Importantly, contact discontinuities do not require regularization via higher-order mechanisms like dispersion or dissipation. Following the KdV and fNLS soliton gas results \cite{carbone_macroscopic_2016, GEl:05} we present here the solution to a Riemann problem for a two-component fNLS breather gas. 

We consider system \eqref{cont}, \eqref{s_12} for $M=2$ and the ``shock tube'' type initial conditions
\be\label{RPf}
\left\{\begin{array}{ll}
w_1(x,0) = w^{1}_0, \ \  w_2(x,0) = 0\, ,     \quad & x<0, \\ [6pt]
w_2 (x,0)= w^{2}_0, \ \ w_1(x,0) = 0,  & x>0,
\end{array}
\right.
\ee
where $w^{1}_0, w^{2}_0 > 0$ are some  constants.  We shall also assume that $s_0^{(1)} > s_0^{(2)} >0$ so that the gases undergo an ``overtaking'' collision resulting in the formation of an expanding ``interaction'' region $c^-t < x < c^+t$ where both component are present. Note that, unlike in the classical gas-dynamics shock tube problem, the initial velocity of the breather gases is not zero but is fully determined, via Eq.~\eqref{s_12}, by the density distribution (\ref{RPf}).

The weak solution for $w_1$ and $w_2$ has a piecewise constant form:
\bea
&&  w_1(x,t)=\left\{\begin{array}{ll}
  w^{1}_0,    \quad   \quad & x\ <\ c^-t, \\ [6pt]
  w^{1}_c, \quad \quad & c^-t\ <\ x\ <\ c^+ t ,\\ [6pt]
 0, \quad \quad   & x\ >\ c^+t.
\end{array}
\right. \nonumber
\\
\label{RPs}
\\
&& w_2(x,t)=\left\{\begin{array}{ll}
  0,    \quad   \quad & x\ <\ c^-t, \\ [6pt]
  w^{2}_c, \quad \quad & c^-t\ <\ x\ <\ c^+ t ,\\ [6pt]
 w^{2}_0, \quad \quad   & x\ >\ c^+t.
\end{array}
\right.
\nonumber
\eea
Here $c^-$ and $c^+$ are the velocities of the left and right discontinuity respectively, and $w^{1}_c$, $w^{2}_c$ and $s^{(1)}_c$, $s^{(2)}_c$ are the densities and velocities of the breather  gas components in the interaction region $x \in [c^-t, \ c^+ t]$. The velocities $s^{(1)}_{c}$ and $s^{(2)}_{c}$ are expressed in terms of $w^{1}_c$, $w^{2}_c$ by relations (\ref{s_12}).
The interaction region densities $w^{1}_c$, $w^{2}_c$ and the contact discontinuities' velocities $c^{\pm}$ are found from the Rankine-Hugoniot conditions:
\begin{equation}
\begin{array}{l}
  -c^-[w^{1}_0-w^{1}_c]\ +\ [w^{1}_0 s_{0}^{(1)}-w^{1}_c s_{c}^{(1)}]\ =\ 0 \, , \\
  -c^-[0-w^{2}_c]\ +\ [0-w^{2}_c s_{c}^{(2)}]\ =\ 0\, ,
\end{array}
\label{j1}
\end{equation}
\begin{equation}
\begin{array}{l}
  -c^+[w^{1}_c - 0]\ +\ [w^{1}_c s_{c}^{(1)} - 0]\ =\ 0 \, , \\
  -c^+[w^{2}_c  - w^{2}_0]\ +\ [w^{2}_c s_{c}^{(2)} -  w^{2}_0 s_{0}^{(2)}]\ =\ 0 \, ,
\end{array}
\label{j2}
\end{equation}
resulting in
\bea
 &&  w^{1}_c\ =\ \frac{w^{1}_0(1-\Delta_{21} w^{2}_0)}{1 - \Delta_{12} \Delta_{2 1}w^{1}_0 w^{2}_0} \, , 
 \nonumber \\
  \label{fc}
 \\
 && w^{2}_c\ =\ \frac{w^{2}_0(1-\Delta_{12} w^{1}_0)}{1- \Delta_{12} \Delta_{21} w^{1}_0 w^{2}_0} \, ,
 \nonumber
\eea
\bea
&&  c^-\ =\ s_0^{(2)}\ -\ \frac{(s_0^{(1)} - s_0^{(2)})\Delta_{12} w^{1}_c}{1- (\Delta_{12} w^{1}_c +\Delta_{21} w^{2}_c)},  \nonumber \\
 \label{cpm}
 \\
 &&  c^+ \ =\ s_0^{(1)}\ +\ \frac{( s_0^{(1)}- s_0^{(2)}) \Delta_{21} w^{2}_c}{1-(\Delta_{12} w^{1}_c + \Delta_{21} w^{2}_c)}\, . \nonumber
 \eea

\bigskip
In conclusion we note that, being an integrable hydrodynamic type system, Eq.~\eqref{cont}, \eqref{s_alg} are amenable to the generalized hodograph transform, enabling in principle the construction of all non-constant smooth solutions \cite{tsarev_poisson_1985, dubrovin_hydrodynamics_1989}.  Indeed, a number of nontrivial exact solutions (such as similarity and quasi-periodic  solutions) were obtained in \cite{el_kinetic_2011}. Their physical interpretation in terms of the fNLS breather and soliton gas is an interesting outstanding problem.

\section{Summary and Outlook}

In this paper we have constructed nonlinear spectral theory of breather and soliton gases in the fNLS equation. This was done by considering a special, infinite-genus thermodynamic limit of finite-gap potentials and of the associated nonlinear modulation equations. 

The core result of the paper is the system of nonlinear dispersion relations \eqref{dr_breather_gen1},
\eqref{dr_breather_gen2} for the  spectral parameters of  breather gas: the density of states $u(\eta)$ and its temporal counterpart $v(\eta)$, as well as the wavenumber $\tilde u (\eta)$ and the frequency $\tilde v (\eta)$ of the carrier wave.  These nonlinear dispersion relations yield the integral equation of state  (Eqs. \eqref{eq_state}, \eqref{eq_state_breather}) connecting the velocity $s(\eta)$ of the quasiparticles (breathers)  in the gas with the gas' spectral density of states $u(\eta)$. The respective relations for soliton gas are obtained collapsing the ``exceptional'' spectral band corresponding to the background Stokes mode in the breather gas.

The equation of state \eqref{eq_state}, together with the transport equation \eqref{transport}  for the slowly varying density of states $u(\eta, x, t)$ form the kinetic equation for  breather  gas.  Combining this equation with 
the ``satellite'' kinetic equation  \eqref{transport_tilde}, \eqref{phase_velocity_BG} for the carrier wave parameters $\tilde u(\eta, x, t)$ and  $\tilde s(\eta, x, t)$, we obtained a full set of  equations characterising the macroscopic spectral dynamics in a spatially non-homogeneous breather  gas. These include the kinetic equation for  soliton gas  \cite{GEl:05},  as a particular case.  Our consideration also includes the bound state (non-propagating) breather and soliton gases not considered previously.  One of the immediate  implications of our analysis is the prediction  of the critical state of breather (soliton) gases, which we term breather (soliton) condensate, and whose properties are fully determined  by the interactions between the quasiparticles  in the gas, while the  individual characterization of these quasiparticles is suppressed. The criticality condition \eqref{crit_u} yields the density of states $u_{\rm c}(\eta)$ in the condensate, and  we present two notable examples where this critical density of states can be found explicitly: the bound state soliton condensate and the so-called ``circular'' soliton condensate with the spectrum located on a circumference in the complex plane.

We now outline some  important directions of future physical and mathematical research suggested by our work. 
\begin{itemize}
\item Statistical characterization of the  nonlinear random wave field $\psi(x,t)$ in  breather and soliton gases: namely, the determination of the probability density function $\mathcal{P} (|\psi|)$, the power spectrum, the correlation function etc.---in terms of the  the spectral density of states $u(\eta)$. This is the subject of an ongoing research, and  the results will be published elsewhere. \GE{}

\item Realization  of breather and soliton gases in numerical simulations and laboratory experiments  and verification of the predictions of the spectral theory developed here.
While the numerical realization of the KdV soliton gas has been reported in a number of works (see e.g. \cite{dutykh_numerical_2014}, \cite{carbone_macroscopic_2016})
the challenge of the modelling of a fNLS soliton gas has been successfully addressed only recently in \cite{gelash_strongly_2018, gelash_bound_2019},  where  dense statistical ensembles of $N$-soliton solutions with large $N$ and random, uniformly distributed  phases of the so-called norming constants were constructed numerically based on a specific implementation of the dressing method.
We note that the above papers employ periodic boundary conditions for the numerical realization of $N$-soliton solutions, i.e. they essentially realize  the  finite-gap solutions of genus $2N$ approximating $N$-soliton solutions.
In particular, the  numerical simulations in \cite{gelash_bound_2019}  have shown that  the bound state soliton condensate described by the `Weyl' distribution of the density of states \eqref{Weyl} represents an accurate model for the developed, nonlinear  stage of spontaneous (noise induced) modulational instability of a plane wave \cite{Agafontsev:15, kraych_statistical_2019}. 
 We can add that our preliminary numerical simulations of  fNLS soliton gas dynamics show a very good agreement with the results obtained in this paper, in particular, with the solution \eqref{j1}, \eqref{j2} of the ``shock-tube'' problem  in Sec.~\ref{sec:multi} and  with the formula \eqref{s-vert0}  for the  velocity of a trial soliton propagating through the bound state soliton condensate. These results will be reported in a separate paper.

\item  Yet another promising line of research, where the developed theory can find applications,  is related  to the rogue wave formation (see, e.g., \cite{gelash_strongly_2018} for recent numerical observations of rogue waves in soliton gas). The relative roles of solitons and breathers in the rogue wave statistics of integrable turbulence  have been discussed in \cite{soto-crespo_integrable_2016} based on the numerical implementation of the traditional IST method. The finite-gap theory has proved  a powerful tool for the  description of rogue waves  (see, e.g., \cite{bertola_rogue_2016, bertola_maximal_2017, grinevich_finite_2018, osborne_breather_2019} and references therein), and the application of the spectral theory of soliton and breather gases, in particular, of solutions to the kinetic equation, could be the next important step in this direction.

\item The subject of this paper is closely related to the study of the semi-classical  limit of the fNLS equation for a class of generic potentials with large ``solitonic content''  of  $\sim 1/\epsilon$ solitons, where $\epsilon \ll 1$ is a  small dispersion parameter. Indeed, the fNLS evolution of such potentials is known to typically lead to the appearance of coherent structures of increasing complexity that can be locally approximated by genus $n$ finite-gap solutions, with $n$ increasing in time (see \cite{kamvissis_semiclassical_2003,  tovbis_semiclassical_2004, lyng_then-soliton_2007, bertola_universality_2013, Jenkins:14, GEl:16} and references therein). Our preliminary considerations indicate that for sufficiently large $t$ (and consequently, large $n$) the semi-classical spectrum of these solutions fits into one of the thermodynamic scaling requirements described in this  paper. Taking into account the effective randomization of  phases,  the large $t$ evolution of semi-classical solutions is expected to provide  the dynamical realization of soliton gas construction described in this paper.  Indeed,  some features of the soliton gas development   from an initial rectangular barrier (box) potential  predicted by the semi-classical analysis in \cite{GEl:16} have been recently observed in an optics experiment  \cite{marcucci_topological_2019}.

\item Related to the previous comment, we mention  the possibility of an alternative  construction of a soliton gas via an appropriate limit  as $n \to \infty$ of $n$-soliton solutions (rather than $n$-gap potentials) of the fNLS equation. Indeed, as we already mentioned, this is the way ($n$-solitons for large $n$) the dense soliton gas has been realised numerically in \cite{gelash_strongly_2018, gelash_bound_2019}. In this connection, an extension to the fNLS equation of the theory of the so-called  primitive potentials originally developed in the context of the KdV equation \cite{dyachenko_primitive_2016} could prove useful (see \cite{girotti_rigorous_2018} for a recent study, where a particular infinite-soliton  solution of the KdV equation was constructed in the form of a primitive potential). Additionally, the Lax-Levermore type approach to the semi-classical limit of $N$-soliton solutions of the fNLS equation (see \cite{ercolani_zero-dispersion_2003} and references therein) could provide a complementary tool for the soliton  gas  description, with possible extension to breather gas.

\item This work is concerned with properties of soliton and breather gases in the {\it focusing} NLS equation. A similar theory can be constructed for the {\it defocusing} NLS (dNLS) equation. The  kinetic equation 
for a gas of {\it dark} (grey) solitons would  necessarily require the presence of a nonzero background and the corresponding spectral theory would represent a ``bidirectional'', counterpart of the KdV soliton gas theory due to the spectrum of the  self-adjoint ZS operator associated with the dNLS equation being located on the real line. Kinetic equations for bidirectional dNLS and shallow water soliton gases are the subject of a separate work \cite{congy_bidirectional_2020}.

\item  The effects of  small perturbations (e.g. dissipation, higher order nonlinearity and dispersion, or a trapping potential) on the   dynamics of ``integrable'' soliton  gases are of great interest for applications. In particular,
the properties of  a rarefied  gas of dark solitons in quasi-1D repulsive Bose-Einstein condensates (BECs) in a harmonic trapping potential were considered in \cite{schmidt_non-thermal_2012}, whereas the methods of the experimental realizaton of soliton gas in trapped BECs were discussed in \cite{hamner_phase_2013}.  Additionally, the dark soliton gas type structures have been observed in the  laminar-turbulent transitions in fiber laser \cite{turitsyna_laminarturbulent_2013}. The examples of  relevant mathematical models  include the higher order NLS equations  (deep water waves, nonlinear optics) and the Gross-Pitaevskii equation  (BECs).  

One of the most physically pertinent questions arising in connection with soliton gas dynamics in perturbed systems  is the one of thermalization  and the associated equipartition of energy. This topic is currently under active investigation in the context of  propagation of weakly nonlinear random waves in  perturbed integrable systems (see e.g. \cite{onorato_route_2015} and references therein). On the other hand, it is known that soliton gases in non-integrable systems can exhibit a peculiar ``soliton attractor'' scenario observed numerically in  \cite{zakharov_soliton_1988}. The development of an analytical approach to the description of soliton or breather gas in the perturbed fNLS equation via the tools of nonlinear spectral theory represents  a major challenge.

\item Finally we mention the intriguing connections of the  spectral dynamics of soliton and breather gases with the  generalized hydrodynamics of  many-body quantum integrable systems, which turn out to be governed by the kinetic equations of the type \eqref{transport}, \eqref{eq_state} \cite{doyon_soliton_2018, doyon_geometric_2018, vu_equations_2019}.  Establishing a  bridge between these two topical areas  is yet another promising avenue for future cross-disciplinary research.
\end{itemize}

Concluding, we expect that our work will stimulate further theoretical and experimental studies of soliton and breather gases in various physical contexts including nonlinear optics, water waves and superfluids as well as in connection with quantum integrable systems.

\begin{acknowledgments}
The work of GE was partially supported by EPSRC grant EP/R00515X/2. The authors thank M. Bertola, E. Blackstone, T. Congy,  A. Kuijlaars,  S. Randoux, G. Roberti, M. Sigal and P. Suret for stimulating discussions.  The authors also thank E. Blackstone for the preparation of some figures.
\end{acknowledgments}

\section*{Appendix: Mathematical Underpinnings}

We now outline the rationale behind the derivation of equations \eqref{kdp} , \eqref{omdq} and \eqref{WFR} using  the Riemann-Hilbert problem approach to finite-gap theory   \cite{deift_riemann-hilbert_1997, bertola_maximal_2017}.

\subsection{Riemann-Hilbert problem}

\begin{figure}
\centering
\includegraphics[width= 1 \linewidth]{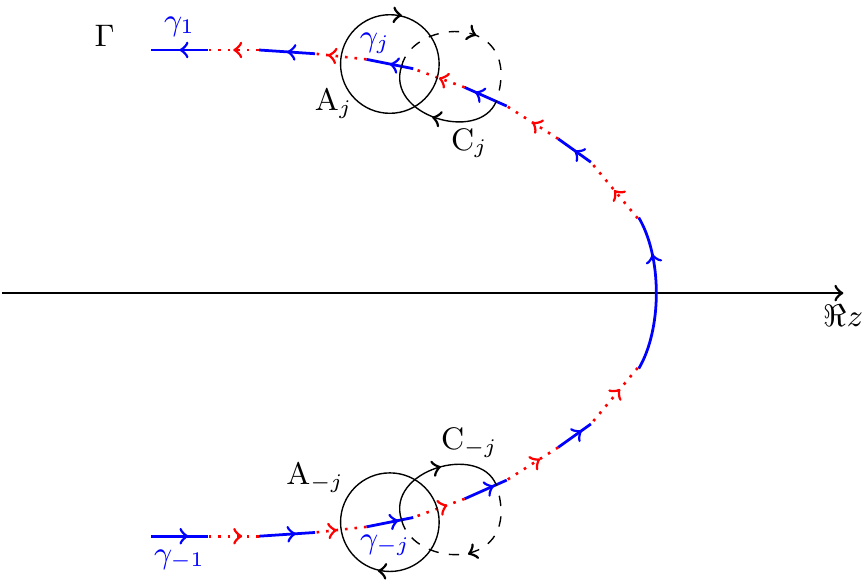}
 \caption{Spectrum geometry for the breather gas construction. The spectral bands (solid lines) lie on the Schwarz-symmetric contour $\G$ in the complex plane $z\in \mathbb{C}$. The cycles ${\mathrm A}_j$ and  $\mathrm C_j$ surround the bands $\g_j$  and
gaps $c_j$ (dotted lines) respectively. {The contours on the second sheet of $\Rscr$ are shown by
 dashed lines.}}
 \label{Fig:Cont2}
\end{figure}

We start with the construction of the real normalized quasimomentum $dp$ and quasienergy $dq$ differentials for the Riemann surface $\Rscr$.
Let $f(z)$ be a polynomial with real coefficients.
Consider the  Riemann-Hilbert Problem (RHP) for the function $g(z)$, that: 
\begin{enumerate}
\item  is analytic in $\C$
with the exception of the jump discontinuity
on all the bands $\g_j$ and all the gaps $c_j$;
 \item 
satisfies  the jump conditions
\bea\label{g-jump-gam}
g_+(z)+g_-(z)&=& f(z) +W_j~~{\rm on}~\g_{ j}, ~~|j|=0,1,\dots, N, \cr~\text{and}\cr
g_+(z)-g_-(z)&=&\O_j~~~{\rm on}~~~c_{ j}, ~~|j|=1,\dots,N, 
\eea
on the bands and gaps, where $ W_0=0$, $W_j=W_{-j}, \, \O_j=\O_{-j}$ and the real constants $W_j, \O_j$, $j=1,\cdots,N$,  are to be determined;  

\medskip and 

\item   is analytic at   $z=\infty$. 
\end{enumerate}
Here $g_\pm(z)$ denote the limiting values of $g(z)$ on the
oriented (see Figure \ref{Fig:Cont2}) bands $\g_j$ and gaps $c_j$. 

By Sokhotsky-Plemelj formula, solution to this RHP is given by 
\begin{equation}\label{gform2}
\begin{split}
g(z)={{R(z)}\over{2\pi i}}
\left[\sum_{|j|=0}^{N} \int_{\g_j}{{[f(\z)+W_j]d\z}\over{(\z-z)R_+(\z)}} \right.  \\ 
\left. +\sum_{|j|=1}^{N} \int_{c_j}{{\O_jd\z}\over{(\z-z)R(\z)}}  \ri].
\end{split}
\end{equation}
Since $g(z)$ must be analytic at infinity, the constants $W_j,\O_j$ must satisfy the linear system 
\be\label{WO}
\begin{split}
\sum_{|j|=1}^N W_j\oint_{A_j}{{\z^m d\z}\over{R(\z)}} +\sum_{|j|=1}^N \O_j\oint_{C_j}{{\z^md\z}\over{R(\z)}} \\
=-\sum_{|j|=0}^N \oint_{A_j}{{f(\z)\z^m d\z}\over{R(\z)}},
\end{split}
\ee
where $m=0,1\cdots, 2N-1$,
obtained from \eqref{gform2} by expanding $\frac{1}{\z-z}$ in powers of $\frac 1z$ as $z\ra\infty$.
{Here $C_j$ denotes the negatively oriented loop around the gap $c_j$, $|j|=1,\dots,N$,  see Fig. \ref{Fig:Cont2}.}

\subsection{Wavenumbers and frequencies}
\label{app:wf}

The system \eqref{WO} for $2N$ real unknowns $W_j,\O_j$ has a unique solution \cite{tovbis_semiclassical_2004} provided that all the branchpoints 
$\a_k$ of the Riemann surface $\Rscr$ are distinct. Thus
\eqref{gform2} indeed defines $g(z)$. 
{Note that $dg$ is a meromorphic differential on 
$\Rscr$ with the only pole at  $z=\infty_2$ on the second sheet of $\Rscr$.
Then   $dh=2dg - df$ is a meromorphic differential on $\Rscr$  
with the  poles at $z=\infty_{1,2}$ on both sheets of $\Rscr$}
 and 
with the periods
\be
\begin{split}
\oint_{\mathrm{A}_{\pm j}}dh = \pm 2(\O_{j-1}-\O_j),  \\ 
\oint_{\mathrm{C}_{\pm j}}dh = \pm 2(W_{j}-W_{j+1}),
\end{split}
\ee
$|j|=1,\cdots, N$, where $\O_0=W_{N+1}=0$. Thus, the differential $dh=d(2g- f)$ is a {second kind} real normalized meromorphic differential with 
the asymptotics $dh\sim - df $ as $z\ra\infty$ on the main sheet.
Since such differentials are uniquely defined by the polynomial $f$ {(\cite{BT15})}, we obtain the 
real normalized quasimomentum $dp$ and quasienergy $dq$ differentials on $\Rscr$ by choosing $f(z)=-z$ and $f(z)=-2z^2$
respectively.
Introducing now
\be\label{k-om}
k_j= 2(\O_j-\O_{j-1}),~~~\tilde k_j= 2W_j,~~~{|j|}=1,\cdots, N
\ee
for $dp,dq$  { with  $f(z)=-z$ and the same expressions for $\o_j,\tilde \o_j$  with  $f(z)=-2z^2$} we obtain  \eqref{kdp} , \eqref{omdq}.
The signs of integrals in  \eqref{kdp} , \eqref{omdq} will be opposite if we replace $j$ by $-j$.

Let us now deform the contours $\g_j$ and $c_j$ of the jump conditions \eqref{g-jump-gam} into $2N+1$ vertical jump contours
 $v_j=[\bar\a_j,\a_j]$, $j=1,\cdots,2N+1$. In particular, $\g_0$ is deformed into $v_1$, the jumps on $c_{\pm N}$ after the deformation become
 distributed between
 $v_1$ and $v_{2N+1}$  etc., see \cite{tovbis_semiclassical_2004}, Section 8, for details. Then we obtain a new RHP with vertical jump contours $v_j$,
 but it is satisfied by the 
  same function $g(z)$ (to be more precise, by the corresponding analytic continuation of $g(z)$). If we would choose the new (deformed)  $\mathrm{A}$-cycles 
  to be the loops around each vertical contour $v_j$, $j=2,\cdots, 2N+1$, then the meromorphic differential  $dh$ would have  all $\mathrm{A}$-cycles
  equal zero, i.e., it would be $\mathrm{A}$-cycles normalized.  
  The wavenumbers and frequences for the fNLS are defined in \cite{dafermos_geometry_1986}, \cite{tovbis_semiclassical_2016} as $\mathrm{B}$ periods of the  $\mathrm{A}$-cycles normalized quasimomentum $dp$ and quasienergy $dq$ differentials
on the Riemann surface of the radical $R(z)$ from \eqref{rsurf} with vertical  branchcuts $v_j$.  
Any two sets of $\mathrm{A}$ and $\mathrm{B}$ cycles of the same Riemann surface are integer linear combinations of each other (see e.g. \cite{dubrovin_theta_1981}). 
As any integer combination of  wavenumbers (frequences)  is also a wavenumber (frequency), we arrive at the statement that $\mathrm{A}$- and $\mathrm{B}$-cycles of $dp, dq$,
defined by \eqref{pq}, are wavenumbers (frequencies) respectively.

\bigskip
\subsection{Nonlinear dispersion relations for finite-gap potentials}
The system \eqref{WO} is the starting point for the derivation of the nonlinear dispersion relations \eqref{WFR}. First of all, introducing 
the constants  $U_k\in \R$ by 
\be\label{def-U}
\O_{l}=\sum_{j=1}^l U_{j} ~~~~~ l=1,\dots ,N,
\ee
we get 
\be\label{Om-U}
\begin{split}
\sum_{|j|=1}^{N} \O_j\oint_{\mathrm{C}_j}{{\z^md\z}\over{R(\z)}}=\sum_{|j|=1}^{N} U_j\oint_{\mathrm{B}_j}{{\z^md\z}\over{R(\z)}}, \\ 
m=0,\dots,2N-1,
\end{split}
\ee
and, thus, reduce  \eqref{WO} to 
\be\label{WU}
\begin{split}
\sum_{|j|=1}^N W_j\oint_{\mathrm{A}_j}{{\z^m d\z}\over{R(\z)}} +\sum_{|j|=1}^{N} U_j\oint_{\mathrm{B}_j}{{\z^md\z}\over{R(\z)}} \\
=- \sum_{|j|=0}^N \oint_{\mathrm{A}_j}{{f(z)\z^m d\z}\over{R(\z)}},
\end{split}
\ee

Taking now linear combinations of equations \eqref{WU}, according to \eqref{Pj}, \eqref{D-P} we obtain
\be\label{WUPj}
W_j+\sum_{|m|=1}^{N} U_m\oint_{\mathrm{B}_m}{{P_j(\z)d\z}\over{R(\z)}}=- \oint_{\gt}{{f(z)P_j(\z) d\z}\over{R(\z)}},~~~
\ee
where  $\gt$ is a large clockwise oriented contour containing all $\g_j$. Taking  imaginary parts of both sides 
of \eqref{WUPj} and substituting $\z$ and $2\z^2$ for $f(\z)$, in view of \eqref{k-om} we obtain \eqref{WFR}.

\subsection{Thermodynamic limit of  nonlinear dispersion relations}
In the thermodynamic limit, the leading order behavior of the coefficients of the linear system \eqref{WUPjM} is given by
\begin{widetext}
\bea\label{ass-coeff}
\oint_{\tilde{\mathrm{ B}}_m}{{P_j(\z)d\z}\over{R(\z)}}=\frac{1}{i\pi}
\le[ \ln\frac{R_0(\eta_j)R_0(\eta_m)+\eta_j\eta_m-\d_0^2}
{R_0(\eta_j)R_0(\bar\eta_m)+\eta_j\bar\eta_m-\d_0^2}
- \ln \frac{\eta_m-\eta_j}{\bar\eta_m-\eta_j}\ri] +O\le( N^2\d^{\frac 23}\ri), \cr
\text{when}~~~m\neq j~~~~\text{and}~~~~~
\oint_{\tilde{\mathrm{ B}}_j}{{P_j(\z)d\z}\over{R(\z)}}=i\frac{2\ln|\d_j|}{\pi}+O(1), 
\eea
 \end{widetext}
where $\tilde{\mathrm{B}}_m =\mathrm{ B}_m + \mathrm{ B}_{-m}$. {Here the branch of $\ln \frac{\eta_m-\eta_j}{\bar\eta_m-\eta_j}$
is defined by the requirement that  $\Im \ln \frac{\eta_m-\eta_j}{\bar\eta_m-\eta_j}=0$ when $\eta_j-\eta_m \in  i\R^+$
and the branch of the remaining logarithm is defined by the requirement that that logarithm is equal zero when $\eta_m\in \R$.}

Substituting these estimates together with \eqref{scal_k_nu} into the  system \eqref{WUPjM} and replacing the Riemann sum with the corresponding integral we obtain \eqref{dr_breather_gen1}, \eqref{dr_breather_gen2}.

%%%%%%%%%%%%%%%%%
%%%%% BIBLIOGRAPHY
%%%%%%%%%%%%%%%%%
%\bibliographystyle{apsrev4-1}
%\bibliographystyle{naturemag}
%\bibliography{breather.bib}

\end{document}